\documentclass[prx,aps,superscriptaddress,reprint,amsmath,amssymb, longbibliography]{revtex4-2}
\usepackage{graphicx}
\usepackage{xcolor}
\definecolor{ultramarine}{RGB}{46,46,145}
\usepackage{amssymb,amsmath,graphicx}
\usepackage{comment}
\usepackage{verbatim}
\usepackage{hyperref}
\hypersetup{
anchorcolor=blue,
colorlinks=true,
citecolor=blue,
urlcolor=blue,
linkcolor=blue}
\usepackage{marginnote}
\usepackage{pdfcomment}

\usepackage{braket}
\usepackage{siunitx}
\usepackage{commath}
\usepackage{textcomp} 

\newcommand{\xtheta}{\hat X_\theta}
\newcommand{\imi}{\mathrm{i}}
\newcommand{\expo}[1]{\mathrm{e}^{#1}}

\newcommand{\wln}{\mathsf{W}}
\newcommand{\ncav}{n_{\rm cav}}
\newcommand{\rhop}{\rho_2^*}

\begin{document}

\title{Wigner negativity in the steady-state output of a Kerr parametric oscillator}

\author{Ingrid Strandberg}
\affiliation{Microtechnology and Nanoscience, MC2, Chalmers University of Technology, SE-412 96
G\"oteborg, Sweden}
\author{G\"oran Johansson}
\affiliation{Microtechnology and Nanoscience, MC2, Chalmers University of Technology, SE-412 96
G\"oteborg, Sweden}
\author{Fernando Quijandr\'{\i}a}
\affiliation{Microtechnology and Nanoscience, MC2, Chalmers University of Technology, SE-412 96
G\"oteborg, Sweden}

\date{\today}

\begin{abstract}




The output field from a continuously driven linear parametric oscillator may exhibit considerably more squeezing than the intracavity field. Inspired by this fact, we explore the nonclassical features of the steady-state output field of a driven nonlinear Kerr parametric oscillator using a temporal wave packet mode description.
Utilizing a new numerical method, we have access to the density matrix of arbitrary wave packet modes. Remarkably, we find that even though the steady-state cavity field is always characterized by a positive Wigner function, the output may exhibit Wigner negativity, depending on the properties of the selected mode.

\end{abstract}

\maketitle

\section{Introduction}\hspace{0pt}

\noindent

As opposed to a quantum field confined in a cavity, a field propagating in free space is characterized by a continuum of modes. In order to give a Schr\"odinger picture description of the propagating output state, wave packet modes are introduced.
In quantum optics experiments, it is possible to infer information about a state stabilized inside a cavity by looking at the output field in a wave packet mode with a bandwidth matching the inverse cavity decay time.
The only requirement imposed on this wave packet function is that it is square-integrable and normalized to unity, which guarantees that the filtered signal corresponds to a single bosonic mode.
Whereas for a linear non-driven system, there exists a unique wave packet mode which maps the cavity state into the output field~\cite{Eichler2011}, this is not the case in the presence of a continuous drive. In this case, nothing restricts us from going beyond the cavity natural bandwidth with arbitrary function profiles.


Depending on the
selected wave packet, properties of the cavity field and its corresponding output field can be quite different. An old and well-known example of this difference is given by the degenerate parametric oscillator (PO). This is commonly employed to generate squeezed states, i.e., states exhibiting quadrature fluctuations below the vacuum level. In the steady-state, the maximum attainable squeezing inside the PO cavity is equal to \SI{50}{\%} of the vacuum noise. Nonetheless, the output field squeezing can largely surpass that, and ideally achieve \SI{100}{\%} squeezing in a narrow frequency bandwidth around the cavity frequency~\cite{Collett1984}.

Squeezed states are a prominent example of nonclassical states of light~\cite{Walls1983Nov,Teich1989Dec,Dodonov2002Jan}. A signature of the nonclassical behavior of a squeezed state is oscillations in the photon-number distribution~\cite{Schleich1987Oct,Zhu1990,Mitra2010Apr}. Such oscillations are not by themselves a signature of a nonclassical state, since classical states can also display population oscillations. A useful nonclassicality criterion which studies the departure of a photon-number distribution from a classical probability distribution was introduced by Klyshko~\cite{Klyshko1996}.
More commonly, the nonclassical nature of a quantum state is defined in terms of so-called quasi-probability distributions~\cite{Cahill1969Jan,Curtright2013Jun}. Examples of these are the Husimi Q-function, the Wigner function and the Glauber-Sudarshan $P$-function. A generally accepted definition of nonclassicality is given in terms of the P-function: if it is negative or more singular than a Dirac $\delta$-function the state is considered nonclassical~\cite{Glauber1963, Sudarshan1963}.
However, the singular nature of a nonclassical $P$-function makes it difficult to access experimentally. Another widely used nonclassicality criterion is negativity of the Wigner function~\cite{Case2008Sep,Kenfack2004}, which in contrast can be directly measured in experiments~\cite{Smithey1993Mar}.
In addition,
negativity of the Wigner function plays an essential role in quantum computation with continuous variables, as it is a distinguishing feature of states that are resourceful for a computational advantage~\cite{Mari2012,Veitch2013Jan}.
In this work we will focus on the Wigner function as an indicator of nonclassical behavior.





The process that generates squeezing in a PO is (degenerate) parametric down-conversion, in which a pump photon of frequency $2\omega$ splits into two photons each with frequency $\omega$. This process occurs in a medium with a second-order nonlinear susceptibility~\cite{Walls}. Real-life nonlinear materials are however not restricted to second order, and higher order nonlinearities may also need to be taken into account. A higher order nonlinearity is also a requirement for the generation of Wigner-negative states. Including a third-order (Kerr) nonlinearity results in the Kerr Parametric Oscillator (KPO) which has recently found applications in microwave quantum optics for the dissipative stabilization~\cite{Touzard2018,Mirrahimi2014Apr,Leghtas2015Feb} and the adiabatic preparation~\cite{Puri2017, Wang2019, Goto2019} of quantum states of light (cat states), which are useful for quantum computing~\cite{Gilchrist2004Jul,Goto2016Feb}. 
Additionally, by utilizing superconducting circuits it is possible to explore the so-called Kerr single-photon regime in which the strength of the nonlinearity surpasses the inverse cavity decay time by an order of magnitude, enabling the observation of previously undetected quantum effects~\cite{Kirchmair2013, Wang2019,Andersen2020Apr,Roberts2020}.

In this work we study the steady-state output from the KPO cavity with respect to temporal wave packet modes. We find that the KPO cavity and output fields can have markedly different nonclassicality properties, mainly in terms of the Wigner function.
For comparison, we also review the PO output field in terms of temporal wave packet modes. While the squeezing spectrum of the output is previously known, we analytically solve for the full output field density matrix, shedding new light on this well-studied system by investigating how its features depend on the properties of the temporal wave packet. Since an analytical solution is not straightforward for the KPO output field, we use numerical simulations using the "input-output with quantum pulses" formalism introduced by Kiilerich and M{\o}lmer~\cite{Kiilerich2020, Kiilerich2020b}. We establish the nonclassical character of the KPO output field through its photon-number distribution using the Klyshko nonclassicality criterion. We find that the presence of the Kerr nonlinearity leads to larger population oscillations than for the PO. We also find that under circumstances that render the steady-state KPO cavity field Wigner-positive, the leaked output field may nonetheless exhibit Wigner negativity, and the magnitude of a Klyshko coefficient directly correlates with the amount of Wigner negativity. Moreover, by numerical optimization we find the temporal wave packet that maximizes the Wigner negativity.

The paper is structured as follows: In Section~\ref{sec:model} we describe our system model. In order to make a comparison with the KPO, we review the PO output field in Section~\ref{sec:PO} before proceeding with the KPO output in Section~\ref{sec:KPO}. There, we compare the population statistics between the cavity steady-state and the output state in order to understand how Wigner negativity can appear in the output despite the cavity being Wigner-positive.  In Section~\ref{sec:KPO_nonclassicality} we take a closer look at the nonclassical properties of the KPO output field for different nonlinearity strengths. Finally, Section~\ref{sec:conclusions} summarizes our results.

\section{The model}\label{sec:model}

Our system is a parametrically driven nonlinear cavity, and we are interested in its steady-state emission. In a frame rotating at the cavity resonance frequency $\omega$, the Hamiltonian of the driven cavity is ($\hbar = 1$)
\begin{equation}\label{eq:Hamiltonian}
\hat{\mathcal{H}} = \frac{1}{2}\left( \beta \hat c^{\dagger 2} + \beta^*  \hat c^2 \right)
+ K \hat c^{\dagger 2 } \hat c^2 ,
\end{equation}
where $\beta$ is the complex two-photon drive (parametric pump) amplitude, $K$ is the Kerr nonlinearity and $\hat c$ ($\hat c^\dagger$) is the annihilation (creation) operator for the photons inside the cavity. This is an ubiquitous quantum optics model which describes squeezing and parametric amplification~\cite{Walls, Gardiner}. The two-photon drive results from the nonlinear interaction of two modes, typically denoted as the signal and the pump. Using the the so-called parametric approximation in which the pump field is assumed to be in a coherent state of large amplitude, i.e.\ classical, the pump mode operator can be substituted by a $c$-number. As an alternative to having a second-order nonlinear crystal in an optical cavity, this Hamiltonian can also be obtained via a time-dependent boundary condition such as a movable mirror~\cite{Bose1997Nov} or a tunable Josephson inductance in a microwave circuit~\cite{Wilson2010,Wustmann2013, Boutin2017}.

The non-unitary dynamics of the cavity is well described by the Lindblad quantum master equation which considers a Markovian (memoryless) environment at zero temperature
\begin{equation}\label{eq:qme}
\partial_t \varrho = -i[ \hat{\mathcal{H}}, \varrho] + \gamma \left( \hat c\, \varrho\, \hat c^\dagger -
 \frac{1}{2} \{ \hat c^\dagger \hat c, \varrho\} \right) ,
\end{equation}
where $\{\cdot,\cdot\}$ denotes the anticommutator and $\gamma$ is the single-photon loss rate of the cavity. In typical quantum optics applications, single-photon loss is the main decay channel of cavity photons into the the continuum of electromagnetic modes which comprise the cavity environment. Here, we consider photon emission into a one-dimensional transmission line. The state of the field $\hat a_{\rm out}$ that leaks out of the cavity can be inferred from the input-output boundary condition~\cite{Gardiner1985}, which relates $\hat a_{\rm out}$ with the drive field and the cavity field operator $\hat c$. Considering the two-photon drive as a weakly coupled input channel, the cavity output field depends solely on the cavity state as
\begin{equation}\label{eq:input_output}
    \hat a_{\rm out}(t) = \sqrt{\gamma}\, \hat c (t).
\end{equation}


\subsection{Wave packet modes}

The cavity output field provides information of the cavity state through the input-output relation~\eqref{eq:input_output}.
However, the cavity field corresponds to a single bosonic mode, while the output
is comprised by a continuum of frequencies. Therefore, in order to do a faithful comparison of the cavity and output states, it is necessary to define a bosonic mode out of this continuum. This is done in terms of wavepacket modes. For convenience, we are going to restrict to a temporal description in the remainder of this work, but the corresponding frequency representation is simply related by a Fourier transform.

We are going to characterize the emission from the cavity in the Fock space defined by the \emph{wavepacket creation operator}
\begin{equation}\label{eq:wavepacket_creationop}
\hat A_f^\dagger = \int_0^\infty {\rm d}t \, f(t)\, \hat a^\dagger_{\rm out} (t ) ,
\end{equation}
with $f(t)$ satisfying the normalization condition
$\int_0^\infty {\rm d}t \, \vert f(t) \vert^2 = 1$ in order
for $\hat A_f$ to fulfill the bosonic commutation relation $[\hat A_f, \hat A_f^\dagger] = 1$.
For simplicity, we restrict $f(t)$ to be a real-valued function. This creation operator defines a
symmetric wave packet in frequency space around the cavity resonance frequency.

Experimentally, detection in a given temporal profile is implemented
by means of a pulsed local oscillator in homodyne detection or by
processing the measurement signal by means of a digital filter
corresponding to the function $f$. For this reason, we
sometimes refer
to the wavepacket profile as a \emph{filter function}.

\section{Parametric oscillator (PO)}\label{sec:PO}

Before moving to the Kerr parametric oscillator, we start by studying the linear, or Gaussian, degenerate parametric oscillator (PO) with $K=0$ in Eq.~\eqref{eq:Hamiltonian}. It is known that the steady-state of the field inside of the resonator exhibits \emph{quadrature squeezing}. This means that the minimum value of the variance of the generalized quadrature operator
$\hat X_\theta = ( \hat b\, {\rm e}^{-\imi \theta} + \hat b^\dagger\, {\rm e}^{\imi \theta} )/ \sqrt{2}$
is smaller than the standard quantum limit set by vacuum noise.
Here $\hat b$ is a generic bosonic field operator $[\hat b, \hat b^\dagger] = 1$, i.e., it can equally refer to the cavity operator $\hat c$ or the filtered output $\hat A_f$. For our chosen normalization of the generalized quadratures, the vacuum noise variance is $1/2$. We label the minimum value of the variance as
\begin{equation}\label{eq:squeezing_s}
    s = \min_\theta \braket{(\Delta \xtheta)^2},
\end{equation}
and consequently, squeezing is indicated by $s < 1/2$. The variance is defined as
\begin{equation}
\braket{(\Delta  \xtheta)^2}=\braket{\xtheta^2} - \braket{\xtheta}^2,
\end{equation}
and it attains its minimum value for a particular value of $\theta$. For our setup, it is dependent on the phase of the drive $\beta$ in Eq.~\eqref{eq:Hamiltonian}. But for simplicity, from now on we are going to restrict to a real-valued two-photon drive amplitude $\beta \in \mathbb{R}$ without loss of generality.

In our particular system we have for the PO steady-state $\langle \hat c \rangle_{\rm ss} = 0$. Consequently, first-moments of the quadrature field also vanish. In general, it is always possible to displace the field in order to meet this condition, so from now on we will ignore first-order moments.
Then, the variance can be rewritten as
$\braket{(\Delta  \xtheta)^2}=  \langle \hat b^\dagger \hat b \rangle
+ \langle \hat b^2 \rangle {\rm e}^{-2 i \theta}/2 + \braket{\hat b^{\dagger 2}}\expo{2\imi\theta}/2  + 1/2$.
Taking into account that $\braket{\hat b^{\dagger 2}}=\braket{\hat b^2}^*$ and noting that $\hat b^\dagger \hat b$ is a positive semi-definite operator, we see that there is squeezing if and only if $\operatorname{Re} [\braket{\hat b^2} \expo{-2\imi\theta}]<-\braket{\hat b^\dagger \hat b}$.
We see that squeezing can only appear if the expectation value $\braket{\hat b^2}$ has a large enough magnitude. This expectation value is associated with off-diagonal terms in the density matrix: coherences between number states $\ket{n}$ and $\ket{n-2}$.

The amount of squeezing of the cavity field increases as we approach the so-called pump threshold $\beta_{\rm th} = \gamma/2$, where the system becomes unstable, i.e., the mean number of photons diverge. However, the maximum attainable squeezing never surpasses \SI{50}{\percent} of the vacuum noise, i.e. $s = 1/4$.
On the other hand, the field leaking out of the cavity can achieve perfect squeezing $s \to 0$ near threshold in a very small bandwidth around the cavity frequency~\cite{Collett1984}. Below we will throw light on this phenomenon in a new way by looking at the filtered output in the Schrödinger picture, demonstrate that the squeezing increases with increased filter time, and provide an intuitive explanation for this. Since an infinite filtering time corresponds to a zero-bandwidth frequency filter at the cavity frequency, our results are consistent with the previous results.

\subsection{PO output state}

With $K=0$, the system is Gaussian since the equation of motion~\eqref{eq:qme} is only quadratic in $\hat c$ and $\hat c^\dagger$.
Hence the output field is completely characterized by its first- and second-order moments for which it is possible to find a closed set of equations. This means we can analytically solve for the state of the output field.
For the selection of a wave packet mode, we choose a constant filter within a time interval $T$, a so-called \emph{boxcar} filter~\cite{Petersen2005, Quijandria2018, Strandberg2019}. This simplifies the analytical calculations.

As already mentioned, we will consider the steady-state emission from the resonator.
Using the quantum regression theorem~\cite{Lax1963, Lax1967, Carmichael1999}
we calculate the steady-state two-time correlations
$\langle \hat c^\dagger (\tau) \hat c(0) \rangle_{\rm ss}$ and
$\langle \hat c(\tau) \hat c(0) \rangle_{\rm ss}$ by assuming the steady-state condition $\partial_t \varrho_{\rm ss} = 0$ in Eq.~\eqref{eq:qme}. Correlations for the filtered output state $\hat A_f$ follow from integration:
\begin{equation}
 \langle \hat A_f^\dagger \hat A_f \rangle = (\gamma / T) \int_0^T {\rm d}t' \, \int_0^T {\rm d}t \, \langle \hat c^\dagger (\tau) \hat c(0) \rangle_{\rm ss}
\end{equation}
and
\begin{equation}
    \langle \hat A_f^2 \rangle = (\gamma / T) \int_0^T {\rm d}t' \, \int_0^T {\rm d}t \, \langle \hat c(\tau) \hat c(0) \rangle_{\rm ss}.
\end{equation}
%
From these correlators we can calculate the covariance matrix elements
$V_{k\ell} = \langle \hat R_k \hat R_\ell + \hat R_\ell \hat R_k \rangle /2 $, for $k,\,\ell=1,\,2$ with
$\hat R_1 = (\hat A_f^\dagger + \hat A_f)/ \sqrt{2}$ and
$\hat R_2 = (\hat A_f^\dagger - \hat A_f)/ \imi \sqrt{2}$,
which completely determine the state of a Gaussian system.
With these matrix elements, the Wigner function can be calculated as
\begin{equation}
    W(x,p) = \frac{1}{2\pi \sqrt{\det\,V}}\exp [-(x,p)^\top V^{-1} (x,p)] ,
\end{equation}
and the Fock space representation of the state is determined
following Refs.~\cite{Dodonov1984,Dodonov1994} (more details in Appendix \ref{app:sect-analytical}).

\begin{figure}[t]
\begin{center}
\includegraphics[width=\columnwidth]{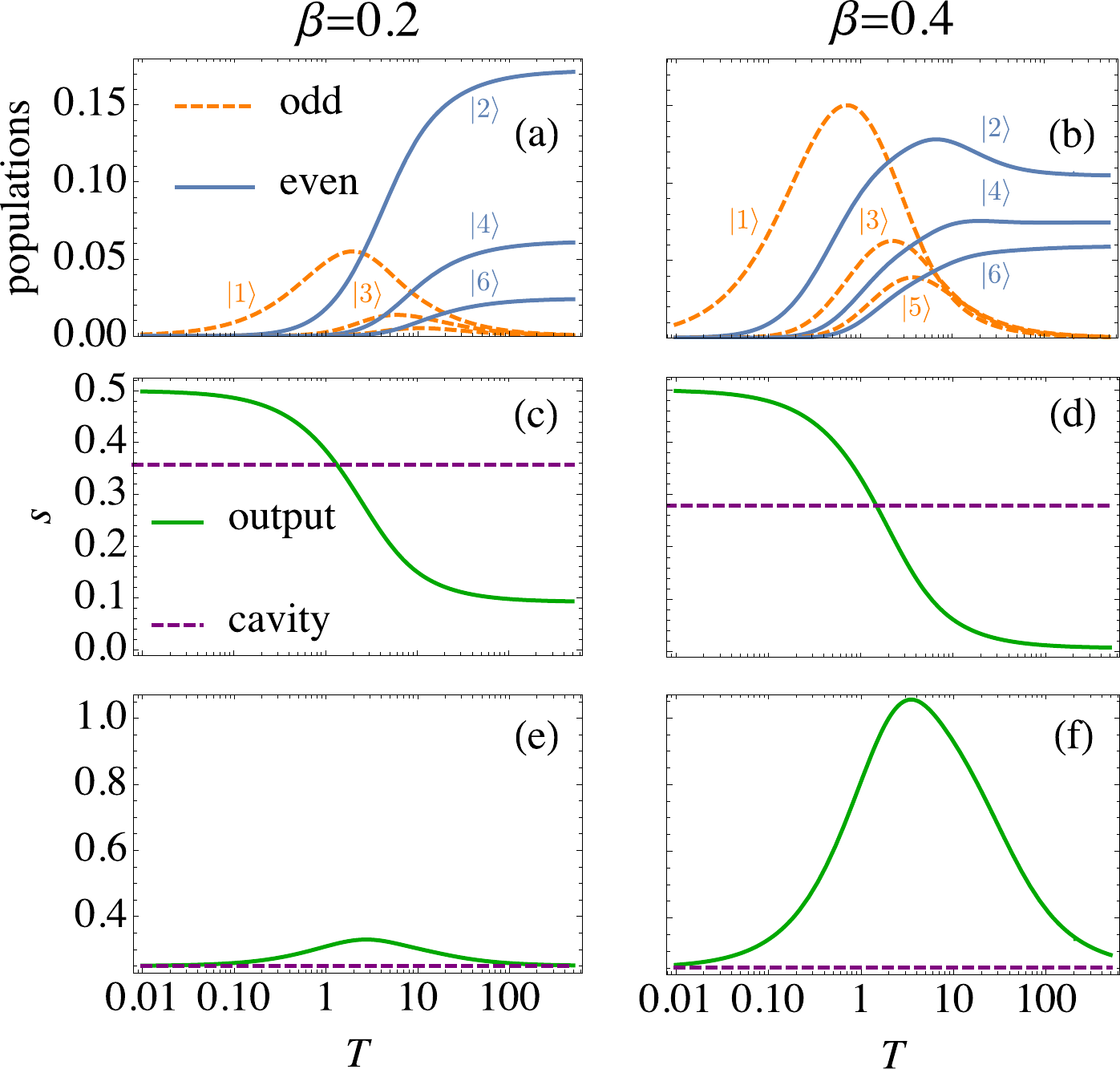}
\end{center}
\caption{Filtered output field populations of the first 6 number states $\vert n \rangle$ excluding the vacuum. Even number states in solid lines and odd number states in dashed lines [(a,b)].
Squeezing for both, the output filtered field in solid lines and the cavity field in dashed lines  [(c,d)].
Determinant of the covariance matrix in solid lines. The dashed lines signal the minimum uncertainty condition [(e,f)]
Everything for $\beta = 0.2$ and $\beta = 0.4$, and as a function of the boxcar field width $T$ in units where $\gamma = 1$.
Note the logarithmic scale on the horizontal axis.
}
\label{fig:grid-PO}
\end{figure}

The behaviour of the filtered output field of the parametric oscillator as a function of the filtering time $T$ for two different drive strengths $\beta=0.2$ and $\beta=0.4$ is shown in Fig.~\ref{fig:grid-PO}. Both $T$ and $\beta$ are given in units where the decay rate is set to $\gamma=1$. In Fig.~\ref{fig:grid-PO}(a) and~\ref{fig:grid-PO}(b) we plot the first six Fock state populations $\rho_n$. For both drive strengths, the single-photon state is the first state to be populated and is the dominant non-vacuum constituent of the complete state for $T \ll 1$.
Progressively, two-, three- and higher photon number states become populated. At around $T \simeq 2$, the two-photon population overcomes the single-photon population.
Increasing the filtering time, we observe the same dynamics for the other even photon number states (the four-photon population overcomes the three-photon one, and so on). As we go towards $T \to \infty$ we see that the odd Fock state populations are suppressed and the state becomes a superposition of even Fock states. From the analytical solution (not shown), we can see that as we approach $\beta_{\rm th}$ in the limit $T \to  \infty$ the populations of all the even states become identical.
This tendency can be seen by comparing the asymptotic populations in\ref{fig:grid-PO}(a) and\ref{fig:grid-PO}(b): the differences between the populations are smaller for the stronger drive.

In Figs.~\ref{fig:grid-PO}(c) and~\ref{fig:grid-PO}(d) we plot the squeezing defined by Eq.~\eqref{eq:squeezing_s} as a function of $T$. Starting as vacuum noise for $T \ll 1$, the output field becomes squeezed as soon as we start to generate photon pairs. It reaches the squeezing level of the intracavity field for $T \simeq 1$, i.e., around the
natural bandwidth defined by the cavity decay rate, and then surpasses it for larger $T$. The maximum squeezing is attained when the state becomes a superposition of even Fock states.

A squeezed state which satisfies the minimum uncertainty allowed by the Heisenberg uncertainty principle is sometimes referred to as an \emph{ideal} squeezed state~\cite{Gerry}. To show that the PO output field becomes an ideal squeezed state in the limit $T \to \infty$ we plot the determinant of the covariance matrix in Fig.~\ref{fig:grid-PO} (e) and (f).
Because it is a real symmetric matrix, the covariance matrix can always be diagonalized.
This diagonal matrix $V$ contains the variances of the squeezed quadrature and the orthogonal one. Their product, i.e., the determinant of $V$, is the uncertainty product.
%
The minimum uncertainty condition in our quadrature normalization corresponds to $\det\,V=0.25$.
For values of $T$ much smaller than the cavity characteristic decay time $1/\gamma$, the minimum uncertainty condition is met because the probability of emission is very small, and thus, we are witnessing the minimum uncertainty of the vacuum state.
As the squeezed states approach the minimum uncertainty condition for large $T$ the populations of the odd Fock states go asymptotically to zero. In Appendix \ref{app:sect-odd-Fock} we show analytically that the odd photon numbers vanish only for a squeezed state which is also a minimum uncertainty state.



\begin{figure}[thb!]
\centering
\includegraphics[width=0.49\columnwidth]{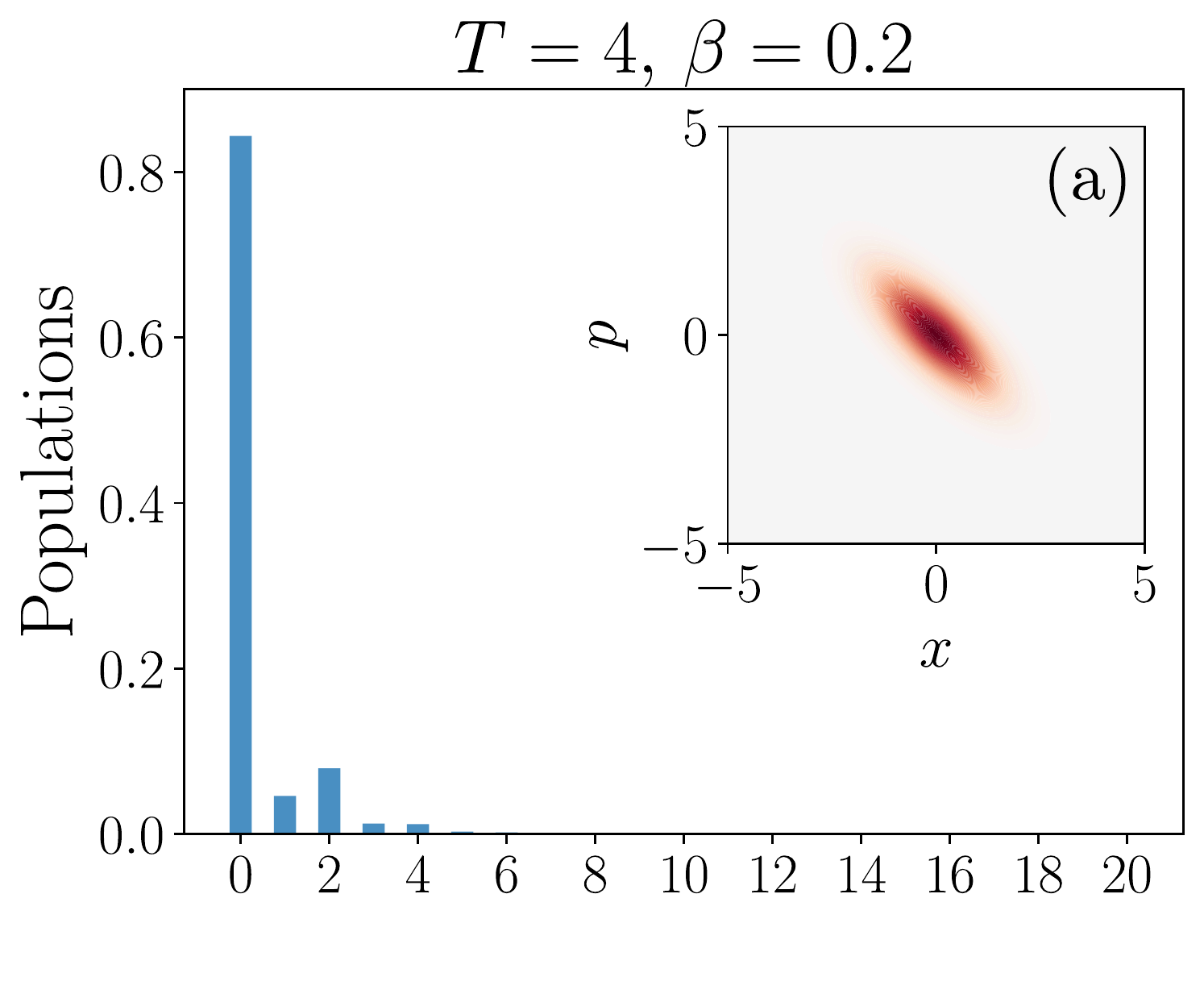}
\includegraphics[width=0.49\columnwidth]{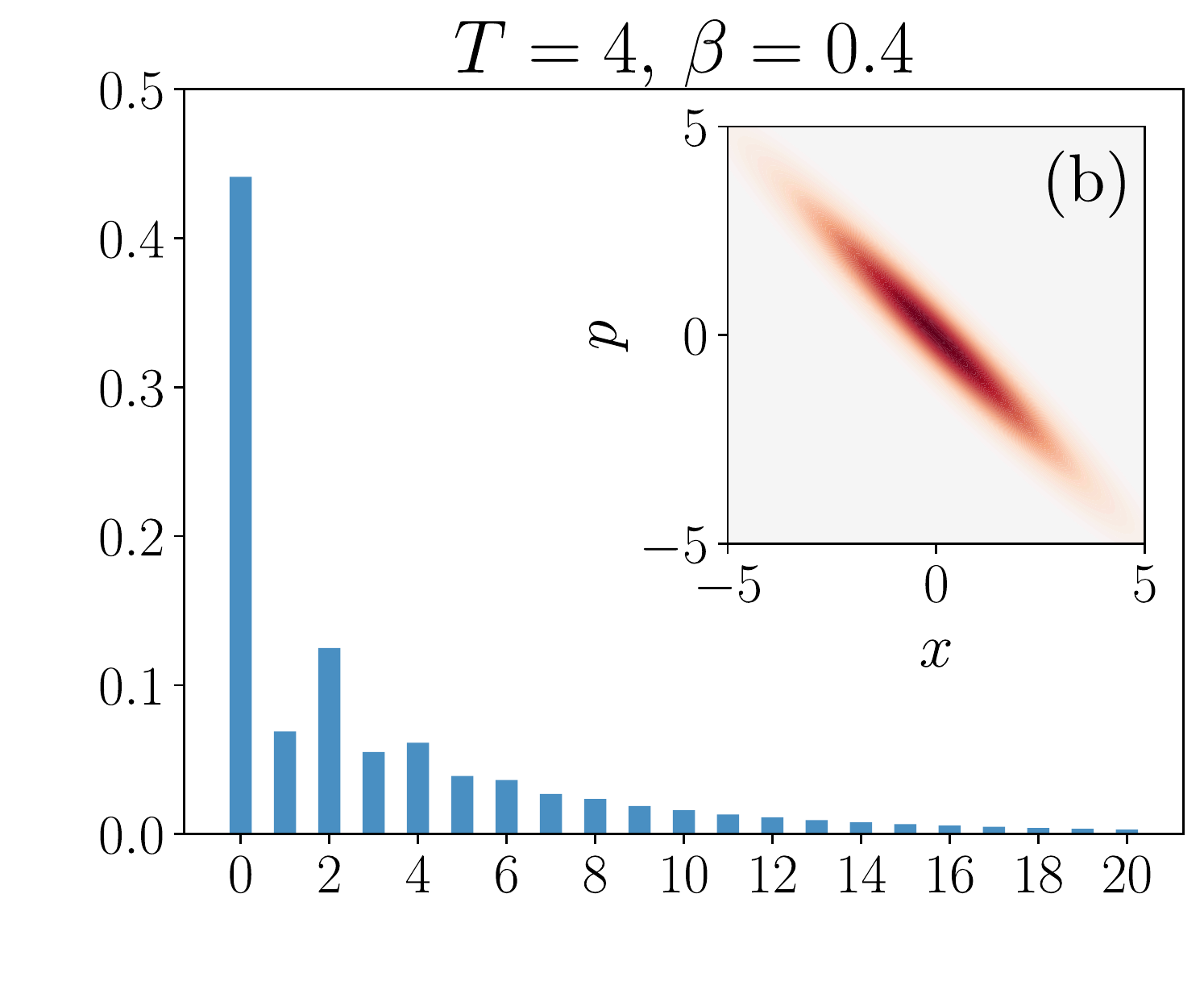}
\includegraphics[width=0.49\columnwidth]{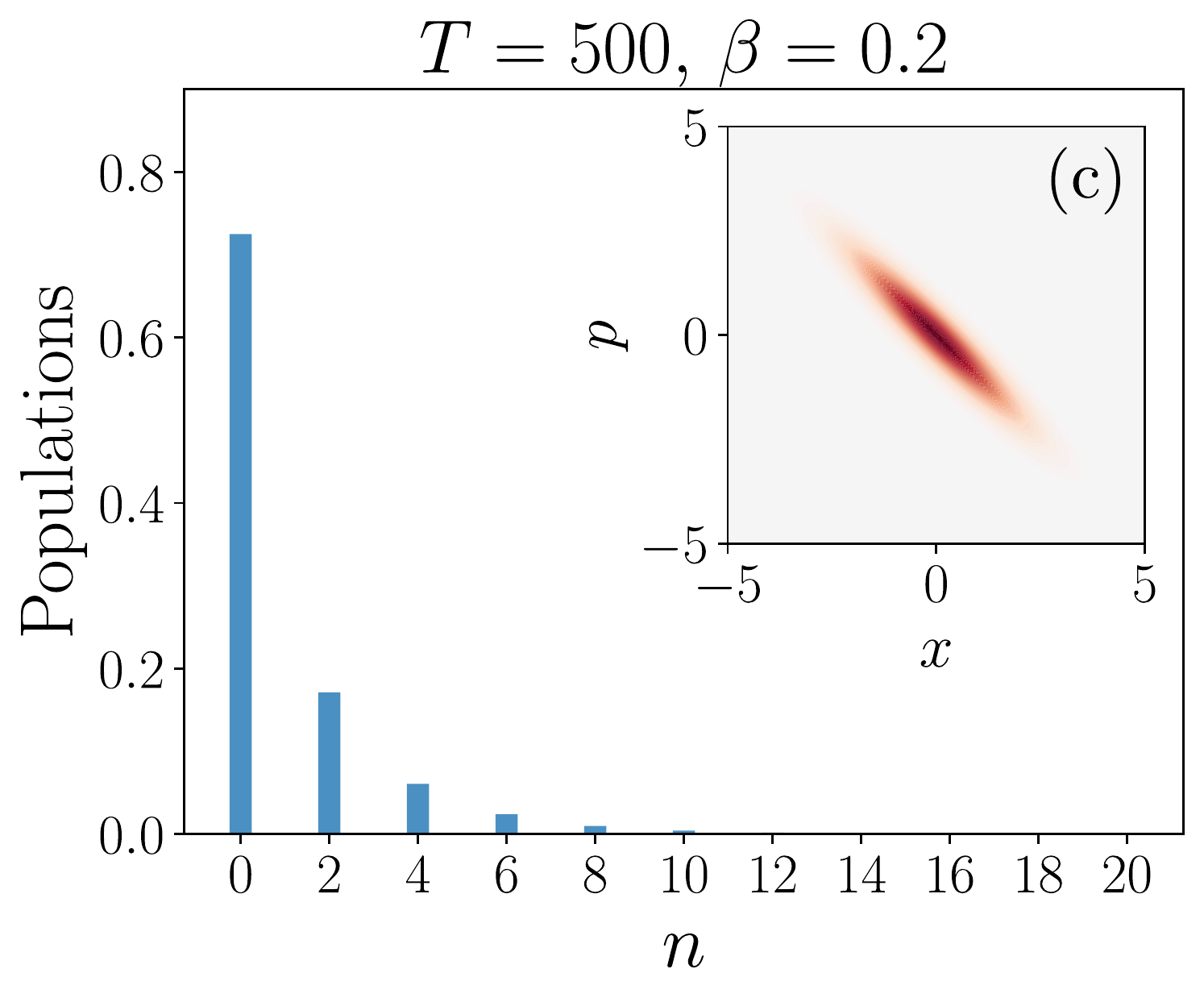}
\includegraphics[width=0.49\columnwidth]{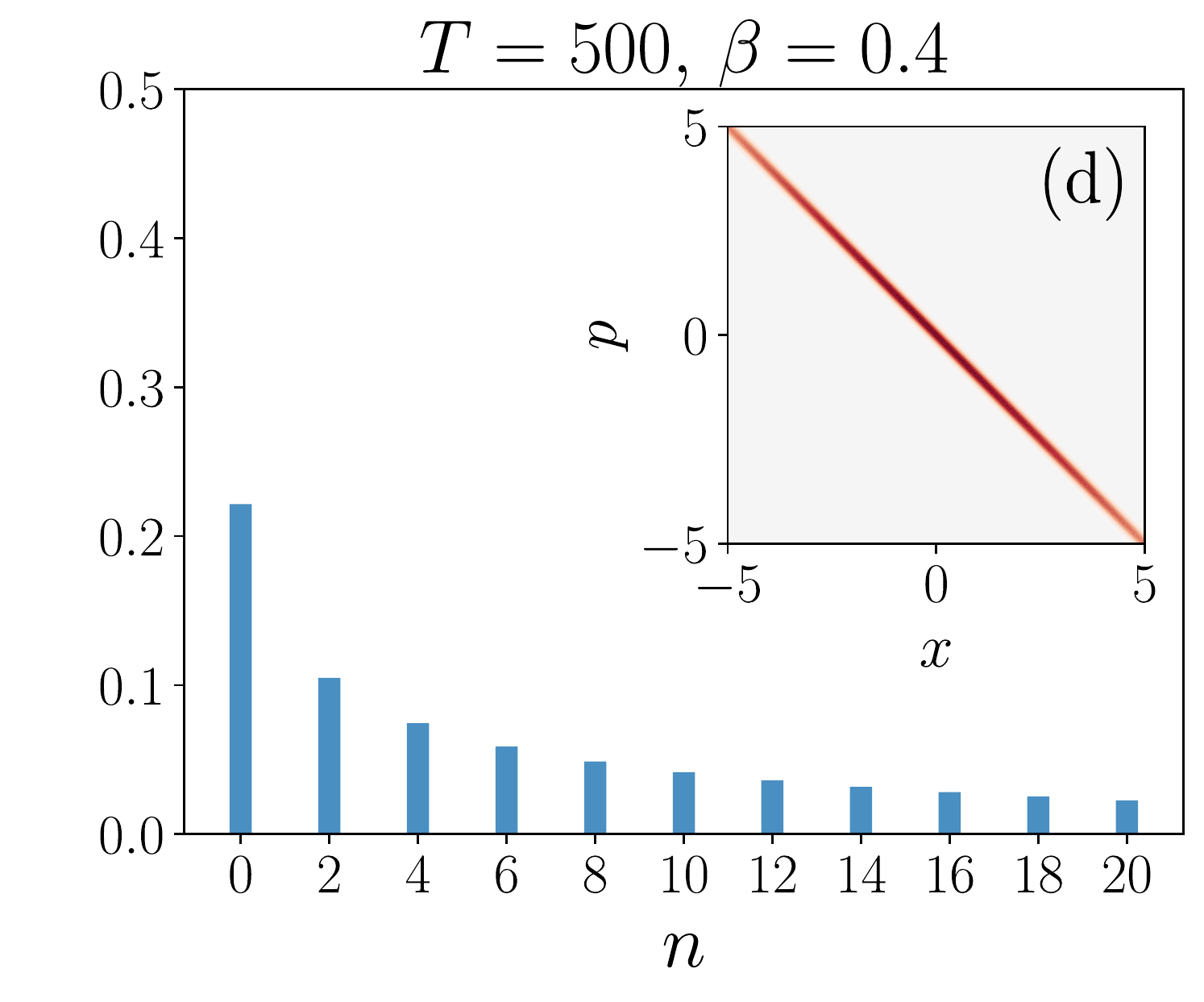}
\caption{PO output field Fock state populations for two different drive strengths $\beta$ and boxcar filter times $T$ (in units where $\gamma = 1$). For the very large filter time $T=500$, only even Fock states are present, and squeezing is enhanced.
}
\label{fig:PO-pops-Wigner}
\end{figure}

Finally, in Fig.~\ref{fig:PO-pops-Wigner} we show the first 20 populations of the output field for two different widths of the boxcar filter, $T = 4$ and $T=500$, for $\beta = 0.2$ as well as $\beta = 0.4$.
We note that vacuum is always the leading contribution to the state.
For $T=4$ we can appreciate an enhanced two-photon population for both drive strengths [Figs.~\ref{fig:PO-pops-Wigner}(a) and~\ref{fig:PO-pops-Wigner}(b)],
although for the stronger drive in~\ref{fig:PO-pops-Wigner}(b) many higher number states are already also populated. For both drive strengths, as $T$ becomes very large, in Figs.~\ref{fig:PO-pops-Wigner}(c) and ~\ref{fig:PO-pops-Wigner}(d) we only observe even Fock states in agreement with Fig.~\ref{fig:grid-PO}. This distinct even-odd oscillation in the populations is a sign of nonclassicality, which will be elaborated further on in section~\ref{sec:klyshko-PO}. The insets in Fig.~\ref{fig:PO-pops-Wigner} show the Wigner functions of the corresponding states, which become more and more squeezed for larger $\beta$ and $T$.


\begin{figure}[t]\centering
\includegraphics[width=1.0\columnwidth]{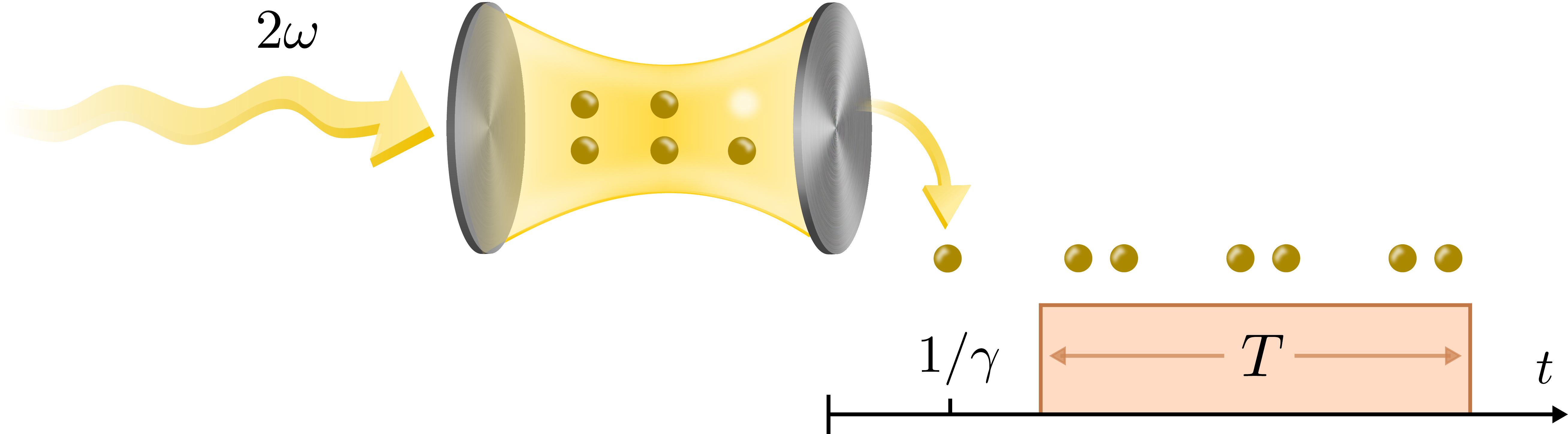}
\caption{A drive photon of frequency $2\omega$ is down-converted into two photons each with frequency $\omega$ in the cavity with resonance frequency $\omega$ and single-photon loss rate $\gamma$. For a longer observation/filtering time $T$, more and more photon pairs are detected.
}
\label{fig:PO1}
\end{figure}

These results can be intuitively understood by means of a very simple time-domain argument as follows: the two-photon drive places correlated pairs of photons inside the resonator. However, the photons leave the cavity one by one. 
This restricts the number of photon pairs inside the resonator and therefore, the maximum amount of squeezing that can be sustained.
On the other hand, we can access the photon pairs by monitoring the output field for a long time compared to the cavity lifetime $1/\gamma$, as illustrated in Fig.~\ref{fig:PO1}. Thus, large two-photon correlations between states $\ket{n}$ and $\ket{n-2}$ can be obtained, and as described in the beginning of this section, therefore more squeezing is expected in the output.

\subsection{Nonclassicality from the PO}\label{sec:klyshko-PO}

There exist several indicators and measures of the nonclassical behaviour of light, including sub-Poissonian statistics~\cite{Davidovich1996}, antibunching~\cite{Paul1982Oct} as well as the previously mentioned photon-number population oscillations and squeezing.
As discussed in the introduction, it is also very common to define nonclassicality by the behavior of quantum phase space quasi-probability distributions such as the $P$-function or Wigner function. 
%
Negativity of the Wigner function
is a sufficient but not a necessary condition for nonclassicality.
For instance, the Wigner functions of the PO cavity and output fields are always positive. This is a consequence of the states resulting from a linear, or Gaussian, system~\cite{Hudson1974}. Nevertheless, quadrature squeezing can be related to a nonclassical $P$-function~\cite{Davidovich1996,Sperling2016Jul}.
Since the squeezed state exhibits population oscillations as the parametric drive creates excitations in pairs, a perhaps more useful~\cite{DAriano1999} nonclassicality criterion for squeezed-like states is given by the Klyshko inequality~\cite{Klyshko1996}
\begin{equation}\label{eq:klyshko}
B_n \equiv (n+1) \rho_{n-1} \rho_{n+1} - n \rho_n^2 < 0 .
\end{equation}
The coefficient $B_n$ is defined in terms of the populations of three consecutive number states $\vert n-1 \rangle$, $\vert n \rangle$ and $\vert n+1 \rangle$ ($n \geq 1$). The Klyshko inequality sets a bound for the population $\rho_n$ in terms of those of its nearest neighbors. If $B_n<0$ for any $n$, the photon-number distribution of the corresponding field departs from a classical probability distribution, i.e., its $P$-function is negative~\cite{Arvind1998}.


The nonclassical nature of states produced in parametric down-conversion has been established via the Klyshko inequality in the optical regime through direct use of photon counters~\cite{Waks2006}. Similarly, in the microwave regime the nonclassical nature of propagating squeezed states generated through a Josephson parametric amplifier was established by feeding the field into a 3D cavity and reading its population content via a dispersively coupled transmon qubit~\cite{Kono2017}.

\begin{figure}[t]
\begin{center}
\includegraphics[width=1.0\columnwidth]{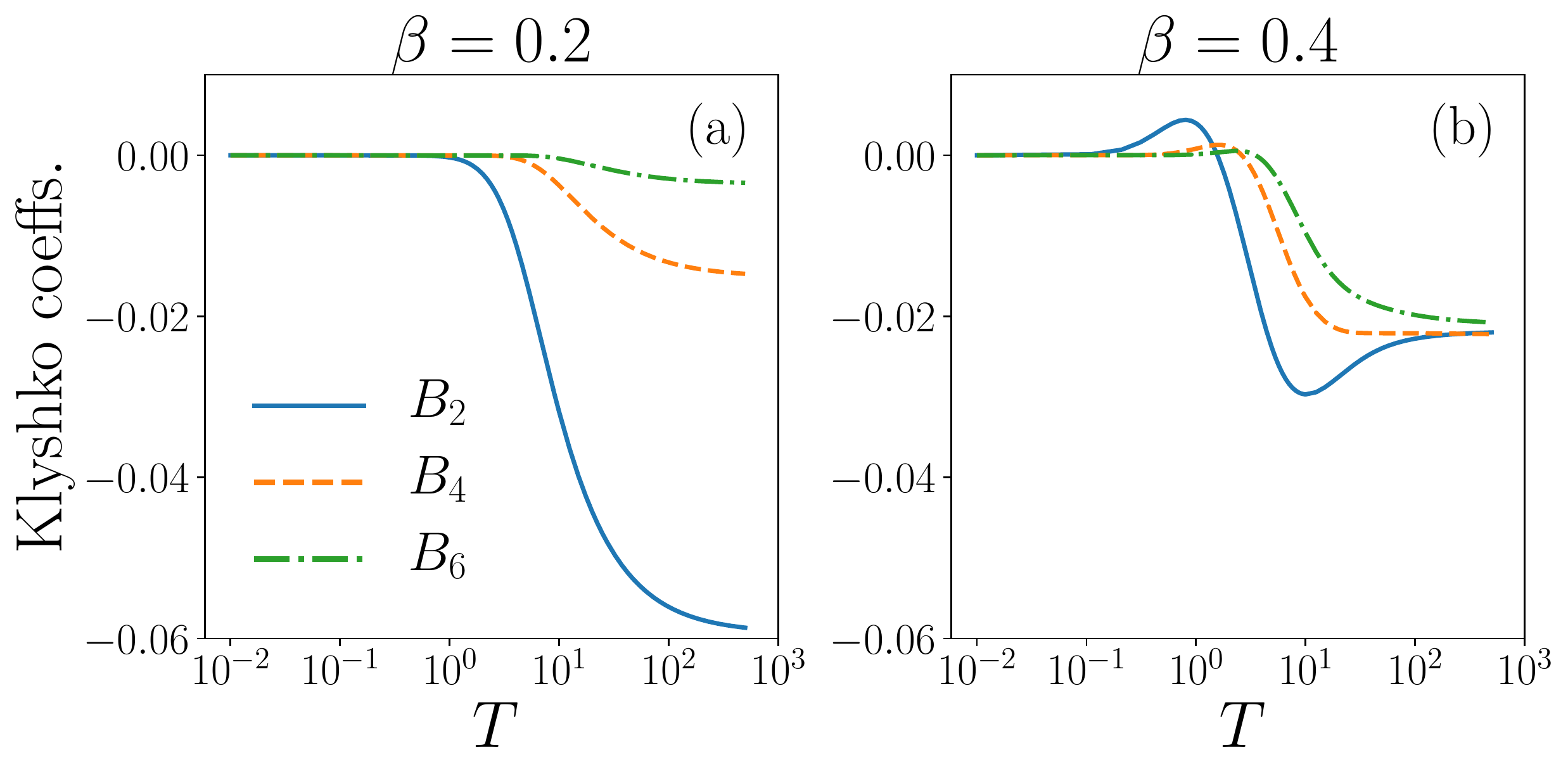}
\end{center}
\caption{ Klyshko coefficients $B_n$
corresponding to the number state $\vert n \rangle$
with $n=2$ (solid line), $n=4$ (dashed line) and $n=6$ (dot-dashed line) for drive strengths
(a) $\beta = 0.2$  and (b) $\beta = 0.4$, for a boxcar filter of width $T$ (in units where $\gamma = 1$). Note  the  logarithmic  scale  on  the  horizontal axis.}
\label{fig:PO-Klyshko}
\end{figure}

The odd Klyshko coefficients are always positive for our system, so in Fig.~\ref{fig:PO-Klyshko} we show the first three even Klyshko coefficients $B_2$, $B_4$ and $B_6$ for the filtered output of the PO as a function of the width $T$ of the boxcar filter function, using two different drive strengths $\beta = 0.2$ and $\beta =0.4$. It can be seen that for $T > 1$ the state is manifestly nonclassical as Eq.~\eqref{eq:klyshko} is first satisfied for $n=2$.
As we have already seen in Figs.~\ref{fig:grid-PO} and~\ref{fig:PO-pops-Wigner}, for $T \gtrsim 2$, the two-photon state becomes the dominant non-vacuum contribution to the output field, and the inequality \eqref{eq:klyshko} for $n=2$ implies that
\begin{equation}
\rho_2 > \sqrt{\frac{3}{2} \rho_1 \, \rho_3}.
\end{equation}
Therefore, when the population of the two-photon state overcomes this bound imposed by the populations of the single- and three-photon states the output field becomes nonclassical. The dominance of $\rho_2$ becomes stronger for larger values of $T$ as the population of the odd number states start to decrease. As $T$ is increased, higher even number states also become populated, which explains the successively appearing negative values of $B_4$ and $B_6$.

Furthermore, we observe a qualitative difference between the behaviour of the Klyshko coefficients for the two drive strengths. While the coefficient $B_2$ exhibits the largest negativity in both cases, for $\beta = 0.2$ it attains its minimum value in the limit $T \to \infty$, when the odd number states are fully suppressed.
On the other hand, for $\beta = 0.4$, as we increase $T$ we involve considerably more (even) number states, each with smaller populations [Cf. Figs.~\ref{fig:PO-pops-Wigner}(c) and~\ref{fig:PO-pops-Wigner}(d)] as $T \to \infty$. The Klyshko $B_2$ coefficient peaks along with the peak of the two-photon population, which is confirmed to occur just before $T=10$ in Fig.~\ref{fig:grid-PO}(b).



\section{Kerr parametric oscillator (KPO)}\label{sec:KPO}


The steady-state of the cavity field for the case of a non-vanishing Kerr nonlinearity has also been exhaustively studied in literature. In fact, despite being a nonlinear system, the steady-state admits an analytical solution. It is a well-established result that the steady-state of our system is characterized by a positive Wigner function.
Specifically, with successively increased drive strength, the cavity steady-state field transitions from vacuum to a weakly squeezed state and finally into an incoherent superposition of coherent states, with quantum coherence washed out by interaction with the environment~\cite{Kryuchkyan1996,Bartolo2016,Roberts2020,Sun2019Sep, Meaney2014}.

While there has been comprehensive studies of the \emph{cavity field}, to the best of our knowledge, a thorough characterization of the properties of the filtered \emph{output field} is still missing in the literature. In the remainder of this paper we are going to study the filtered output field of the KPO, with special emphasis on its nonclassical properties using the above introduced Klyshko nonclassicality criterion as well as negativity of the Wigner function.

Characterizing the state of the propagating cavity output field is, in general, a difficult problem since it implies the calculation of multi-time field correlations for different time orderings.
For a Gaussian or linear system like the PO studied in Sect.~\ref{sec:PO}, the output field is completely characterized by its first- and second-order moments for which it is possible to find a closed set of equations. In the presence of the nonlinearity, this is no longer possible. Instead, we would get an infinite number of equations involving every possible order of the output field moments. For this reason we do not attempt to characterize the output field by calculating multi-time correlations, but instead follow a different approach. In order to explore the features of the output field of the KPO we are going to make use of a technique recently introduced by Kiilerich and M{\o}lmer~\cite{Kiilerich2020, Kiilerich2020b}. We implemented it using QuTiP~\cite{Qutip2013}, and the code can be found in Ref.~\cite{code}. The numerical solutions were validated against the analytical solution for the PO. Alternatively, one could rely on stochastic methods to mimic a quantum tomography experiment, as done in Refs~\cite{Quijandria2018, Strandberg2019}, or numerically solve the full Schr\"odinger equation for the coupled cavity and output fields~\cite{Goto2019}.

Before discussing the nonclassicality of the KPO output in depth, we will introduce the Poissonian regime of the KPO, in which both the cavity and the output are classical states with Poissonian photon-number statistics.
We then then discuss the two-photon population of the KPO output, which will be shown to have a strong connection to its nonclassical features.

\subsection{KPO cavity Poissonian regime}\label{sec:KPO_cavity_poissonian}

The nonlinearity prevents the instability at the drive threshold $\beta = \beta_{\rm th}$. This means that the mean number of photons in the cavity $\ncav$ no longer diverges at this point, but grows steadily with $\beta$. When the photon number reaches $n_{\rm cav} \gg \gamma/ 8K$, the cavity steady-state is an incoherent superposition of coherent states~\cite{Puri2017}. Here the photon-number
distribution is Poissonian, $\rho_n = n_{\rm cav}^n \exp(-n_{\rm cav})/ n!$. Therefore, we will refer to this type of field configuration as the \emph{Poissonian regime}.


The maximum of a Poissonian photon-number distribution is centered around the average number of photons present in the field. So while the \emph{average} number of photons in the cavity grows along with the parametric drive strength $\beta$, when the Poissonian regime is reached, individual number state populations reach their peak and start decreasing in succession as the peak of the Poissonian distribution shifts to higher and higher number states.
As such, we expect that the largest achievable population for the $n$th number state happens when the cavity field populations has a Poissonian distribution with $\ncav= n$.
Then, it is possible to estimate the largest two-photon population in the KPO cavity, which is $\rho_2^* \simeq 0.27$.

In the next section we study how the KPO output field two-photon population depends on the filtering time and the nonlinearity strength. Similarly to the PO, we can get a two-photon population that dominates over the one- and three-photon populations for filtering times comparable to or longer than the cavity decay time.

\subsection{Filtered KPO output field---boundary of the Poissonian regime}

Intuitively, the output two-photon population is expected to grow with $\beta$ until the system reaches the Poissonian regime, similarly to the cavity state.
Here we are going to show that, in contrast to the cavity field, for the output field it is possible to produce states which exhibit the same maximum two-photon state population $\rho_2^*$ but with non-Poissonian photon number statistics.
This is significant, because in Appendix~\ref{app:sect-coherent} we show that when the cavity is in the Poissonian regime, so is the output field.
Specifically, the output field is also a classical mixture of coherent states, with the average number of photons $N_f$ given by the cavity photon number $\ncav$ and a scale factor depending on the width of the filter function. %

For a boxcar filter the mean number of photons in the output field is given by $N_f = \gamma \ncav T$.
Thus, in the Poissonian regime, the largest two-photon population is achieved when $N_f = 2$. Then, from the photon-number scaling relation, we can calculate the time $T^*$ at which $N_f = 2$.
%
This corresponds to $T^* = 2 / ( \gamma \, n_{\rm cav})$~\footnote{
Above threshold and in the Poissonian regime, we have
$\ncav \approx (1/2)\sqrt{(\beta/ K)^2 - 0.25(\gamma/K)^2}$,
which leads to~\cite{Puri2017}:
$ T^* \approx 4/(\gamma \, \sqrt{(\beta/ K)^2 - 0.25(\gamma/K)^2})$.
}.
Therefore, the largest two-photon population
expected in the output of the KPO in the Poissonian regime is $\rhop\simeq 0.27$ for a filtering time $T^*$.

We need to be outside of the Poissonian regime to observe nonclassical effects. Below, we show the behavior of the photon populations in the filtered output field as we depart from this regime for different values of the nonlinearity strength.
As stated in section~\ref{sec:KPO_cavity_poissonian}, not being in the Poissonian regime implies a limited drive strength $\beta$, which in turn requires an increased filtering time $T$ in order to collect a significant number of photon pairs.

\begin{figure}[hbt!]
\centering
\includegraphics[width=0.8\columnwidth]{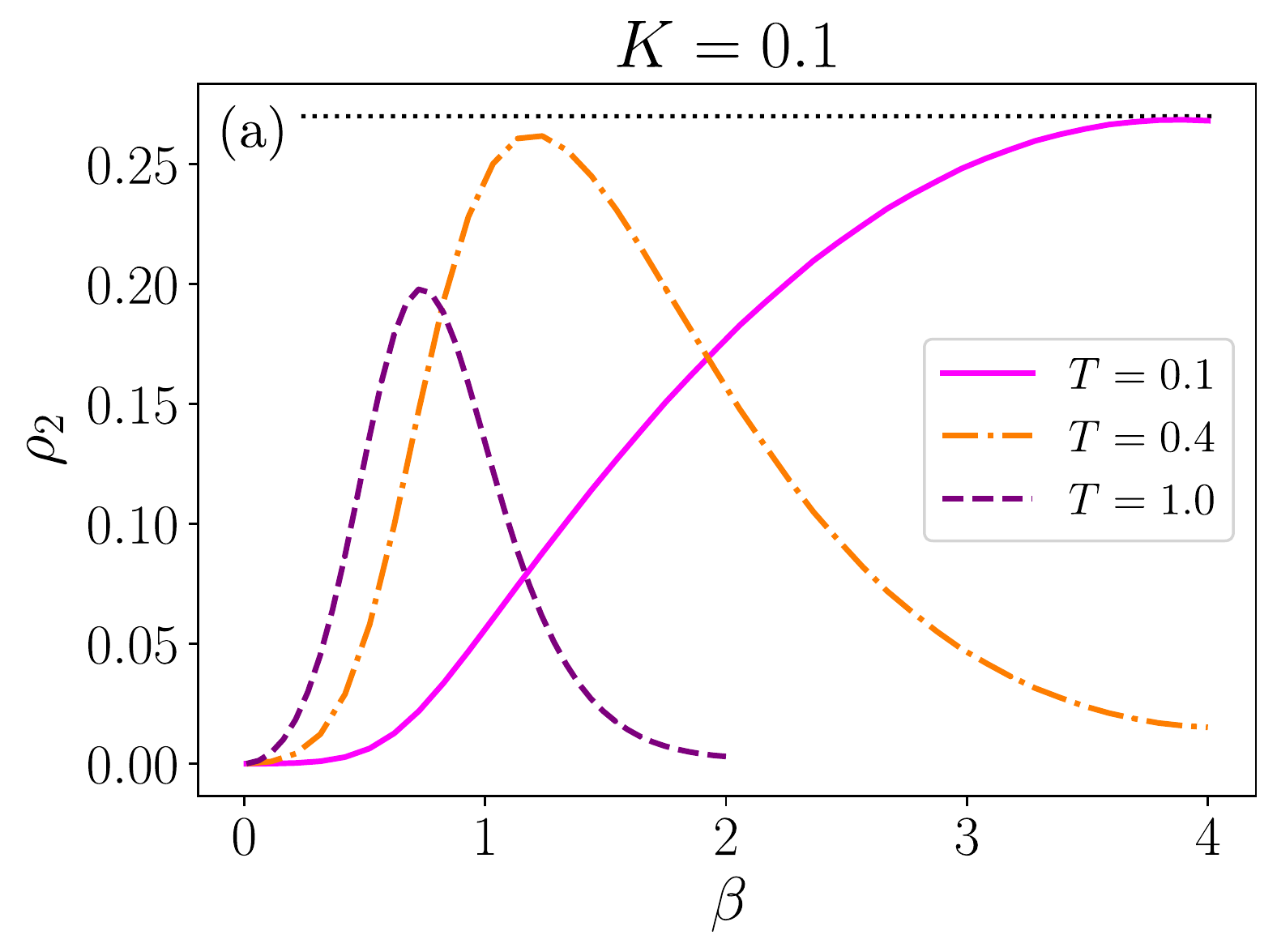}
\includegraphics[width=\columnwidth]{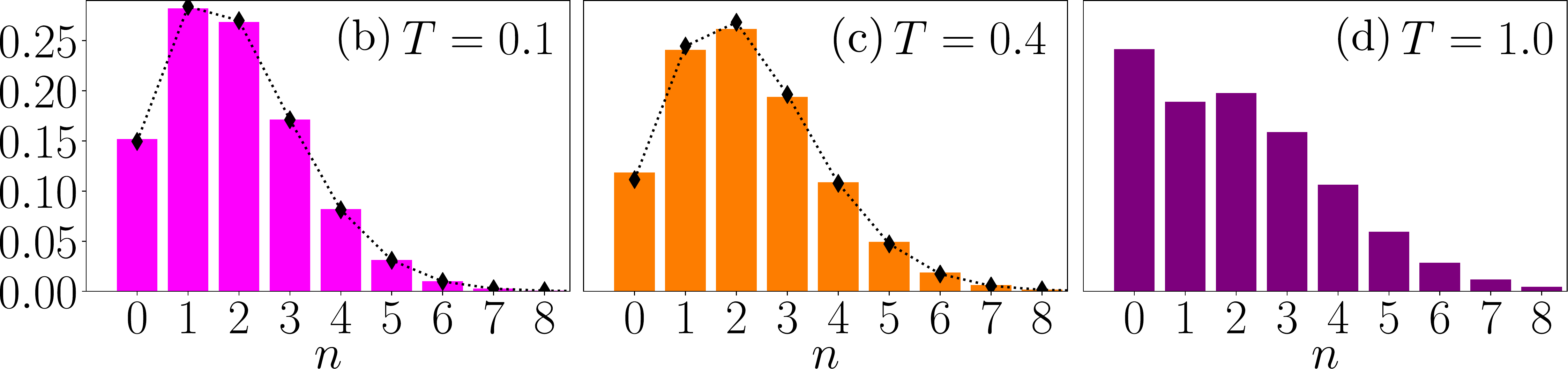}
\caption{(a) Output two-photon population for $K=0.1$ as a function of $\beta$ for $T = 0.1$, $0.5$ and $1.0$ in units where $\gamma = 1$. The dotted line at $\rho_2=0.27$ represents the theoretical maximum two-photon population in the Poissonian regime. Note that the two-photon population is decreased when moving out of the Poissonian regime. Photon number distributions at the two-photon peaks for (b) $T=0.1$ fitted to a Poissonian distribution with $\ncav=1.9$. (c) $T=0.5$ fitted to a Poissonian distribution with $\ncav=2.2$. (d ) $T=1.0$ does not fit a Poissonian distribution.}
\label{fig:weak-K}
\end{figure}

In Fig.~\ref{fig:weak-K}(a) we show the output two-photon population as a function of $\beta$ for a weak nonlinearity $K=0.1$ and different values of $T$ (in units of $\gamma=1$). As expected, we observe that the two-photon population peaks at weaker drive strengths for larger values of $T$, and vice versa.
Additionally, we note that as $T$ is increased, the photon population statistics change. For the shortest filtering time $T=0.1$, the drive strength $\beta$ at which the maximum two-photon population is attained corresponds to the cavity field being in the Poissonian regime, and subsequently, the output is also Poissonian, as can be seen from the photon-number distribution in~\ref{fig:weak-K}(b). Here the largest two-photon population observed corresponds to $\rhop \simeq 0.27$ (dotted line) and the filtering time at which it is achieved agrees with $T^*$.
For a slightly increased $T$ it is possible to generate output states with a Poissonian-like photon-number distribution [cf.\ \ref{fig:weak-K}(c)] but with a smaller two-photon content than $\rhop$, as for $T=0.4$. The photon-number distribution of the output state slowly departs from the Poissonian behavior as we further increase $T$, as exemplified with $T=1$ in Fig.~\ref{fig:weak-K}(d). But crucially, the two-photon population is reduced more and more as $T$ is increased and we move further out of the Poissonian regime.

\begin{figure}[hbt!]
\centering
\includegraphics[width=0.8\columnwidth]{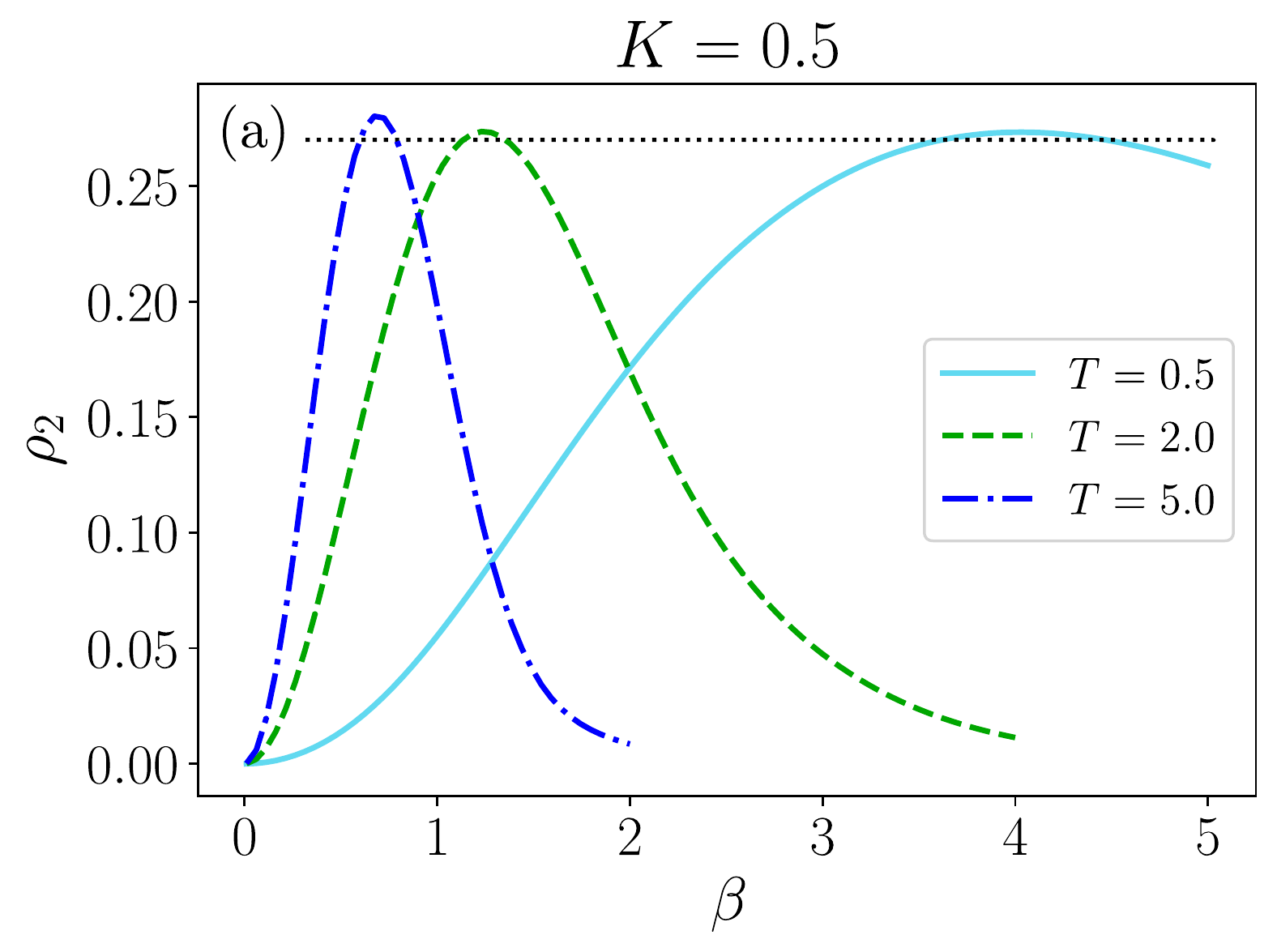}
\includegraphics[width=\columnwidth]{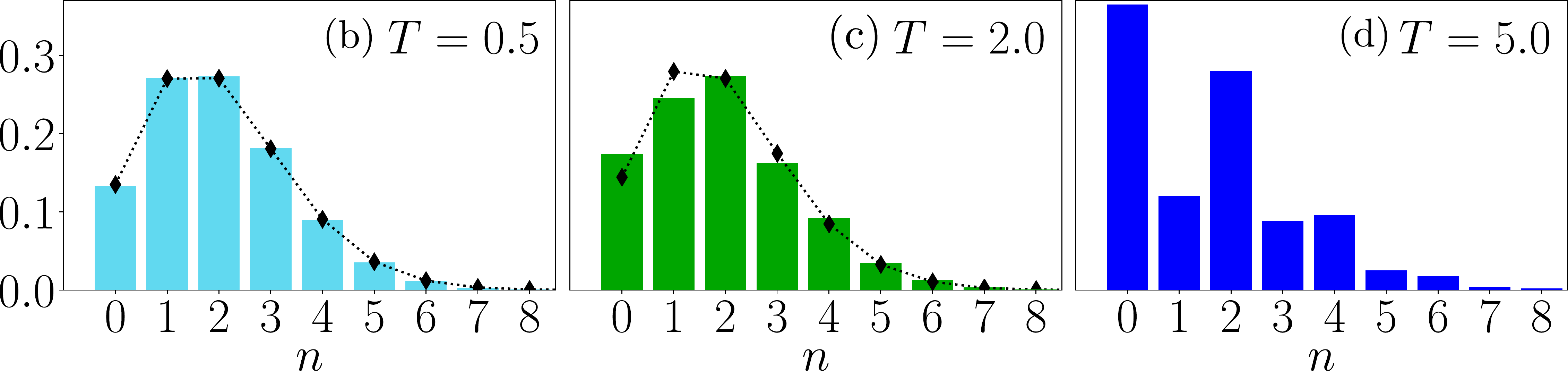}
\caption{(a) Output two-photon population for $K=0.5$ as a function of $\beta$ for $T = 0.5$, $2.0$ and $5.0$ in units where $\gamma = 1$. The dotted line at $\rho_2=0.27$ represents the theoretical maximum two-photon population in the Poissonian regime. The two-photon populations are close to the maximum both within and outside of the Poissonian regime (in stark contrast to the behavior in Fig~\ref{fig:weak-K}) with a maximum if $\rho_2=0.28$ obtained for the most non-Poissonian state corresponding to $T=5.0$. Photon number distributions at the two-photon peaks for (b) $T=0.5$ fitted to a Poissonian distribution with $\ncav=2.0$. (c) $T=2.0$ does not quite fit a Poissonian distribution. (d) $T=5.0$ shows clearly non-Poissonian statistics with population oscillations.}
\label{fig:strong-K}
\end{figure}

Interestingly, we do not need a strong nonlinearity to get dramatically different behavior. In Fig.~\ref{fig:strong-K} we show results for a moderate nonlinearity $K=0.5$. Here, as we decrease the drive strength $\beta$ and increase the filtering time $T$, we depart from the Poissonian regime similarly as with the weak nonlinearity. But in contrast, even when not in the Poissonian regime, the largest two-photon populations for $K=0.5$ still correspond approximately to $\rhop$. In fact, from our numerical simulations the largest two-photon output even can even slightly exceed this value, as can be seen in Fig~\ref{fig:strong-K}(a).

%
For the longer filtering time in Fig~\ref{fig:strong-K}(d), population oscillations start to become evident. But it can be noted that opposed to the PO, for which we know that the odd number states can be completely suppressed for a long filtering time as shown in Sect.~\ref{sec:klyshko-PO}, it is not possible 
with a nonzero Kerr nonlinearity [c.f. Appendix~\ref{app:sect-odd-Fock}].

In the next section we are going to study the nonclassical features of the filtered output field of the KPO.
As described in section~\ref{sec:klyshko-PO}, nonclassicality is not only a consequence of the enhanced two-photon population, but by how large it is compared to its nearest neighbors, i.e., single- and three-photon states. For example, the photon distributions in Figs.~\ref{fig:strong-K}(b),~\ref{fig:strong-K}(c) and~\ref{fig:strong-K}(d) have roughly the same two-photon populations, yet only~\ref{fig:strong-K}(b) corresponds to a Poissonian distribution. In~\ref{fig:strong-K}(c), the state begins to depart from the Poissonian regime and the two-photon population overcomes that of its nearest neighbors. Further away from the Poissonian regime this asymmetry is even larger, as in~\ref{fig:strong-K}(d). We are going to quantify this using the Klyshko coefficient $B_2$.

\section{Nonclassicality in the KPO output}\label{sec:KPO_nonclassicality}

\begin{figure}[hbt!]
\centering
\includegraphics[width=0.8\columnwidth]{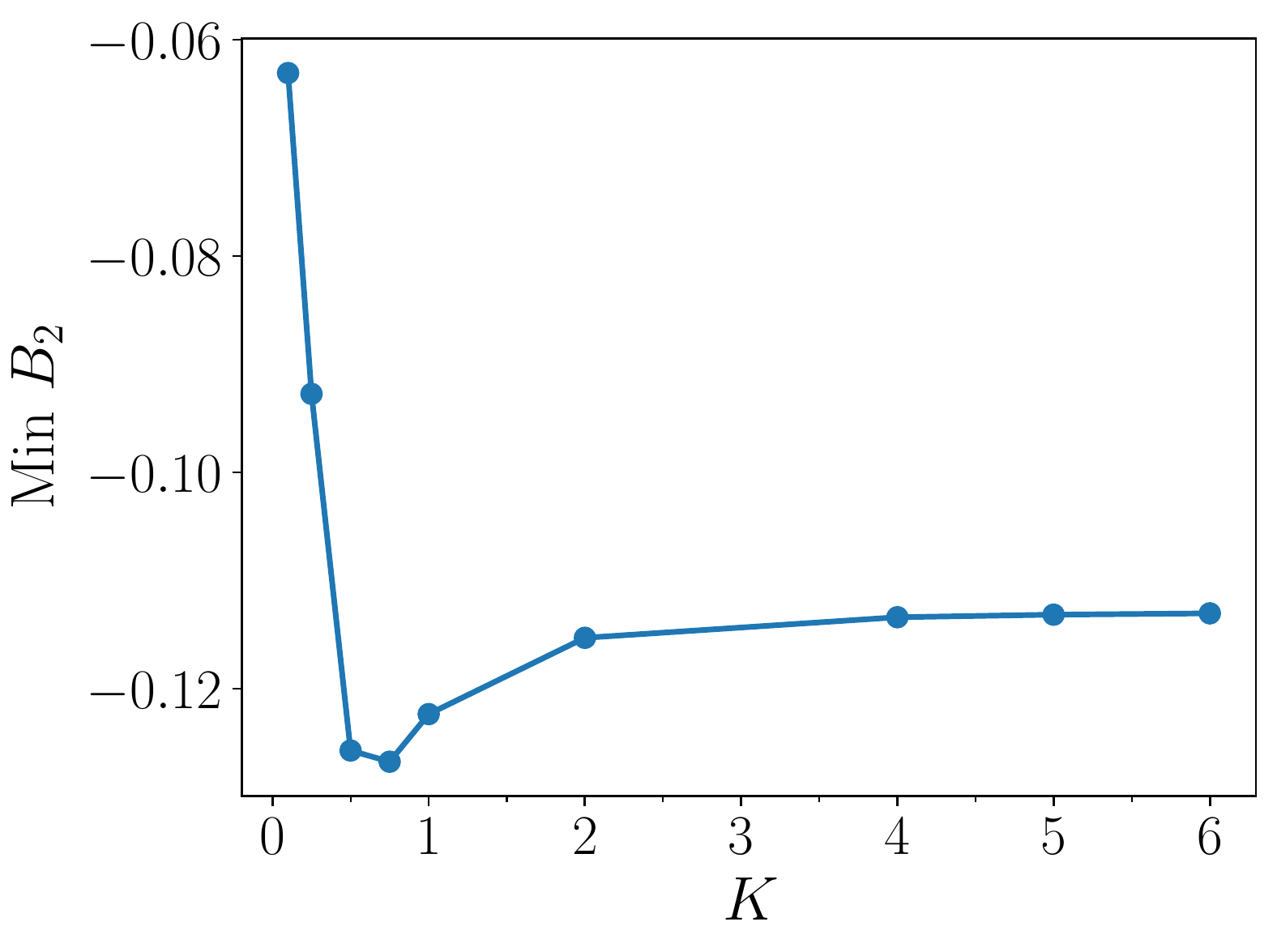}
\caption{ 
When the nonlinearity $K$ is increased from zero, the Klyshko $B_2$ coefficient rapidly drops to its minimum for $K=0.7$, corresponding to the "most nonclassical" state as determined by the Klyshko criterion. As $K$ is further increased, the value of $B_2$ increases slightly before saturating.
The minimum values were obtained by sweeping over a grid of $\beta$ and $T$ for each $K$.
}
\label{fig:klyshko_K_min}
\end{figure}

In this section, we will evaluate the nonclassicality of the KPO output field in terms of the Klyshko $B_2$ coefficient and Wigner negativity.
%
Starting with the Klyshko criterion, we display the minimum $B_2$ for different nonlinearities $K$ in Fig.~\ref{fig:klyshko_K_min}. The Klyshko inequality $B_2 < 0$ is satisfied which establishes the nonclassical nature of the KPO output field. Specifically, when $K$ grows toward 0.5 and larger, the enhanced two-photon contribution leads to a very sharp population oscillation [an example of this is Fig.~\ref{fig:strong-K}(d)] and consequently, to a very large magnitude of the Klyshko coefficient.
The largest magnitude is obtained for $K=0.7$ A stronger nonlinearity does not increase the nonclassicality, in fact, it decreases slightly before it saturates for $K\gtrsim 2$. Further increasing $K$ does not lead to fundamentally different behavior, since the cavity steady-state field is entirely determined by the ratio $\beta/K$ in the strong nonlinearity regime~\cite{Kryuchkyan1996, Meaney2014, Bartolo2016, Roberts2020}. For a weak, or even vanishing, nonlinearity the value of $B_2$ can remain $<0$, but the magnitude is severely diminished compared to what is attainable with a moderate or large $K$.




\begin{figure}[ht!]\centering
\includegraphics[width=\columnwidth]{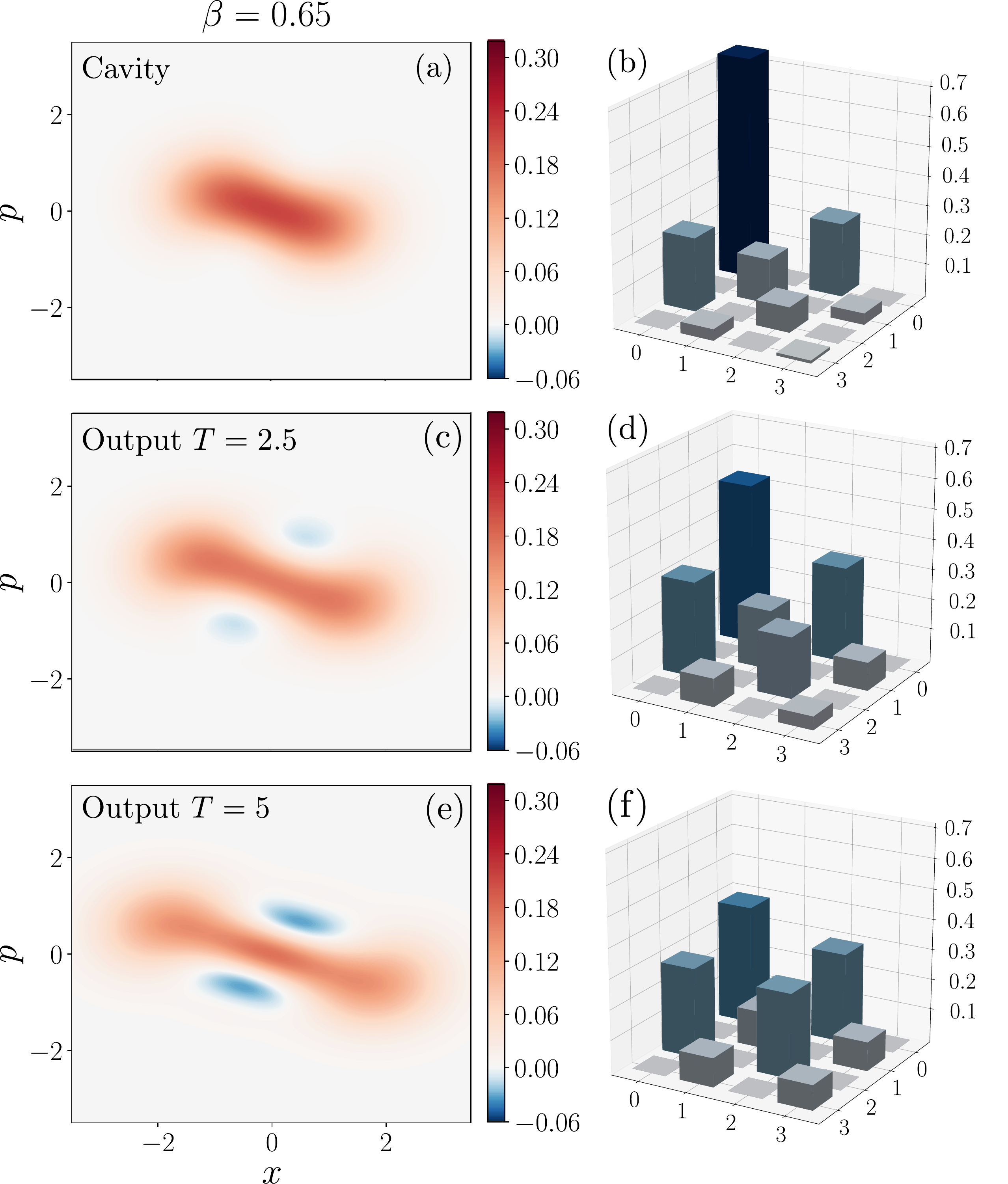}
\caption{Wigner functions and density matrices for $\beta=0.65$. (a) Wigner function and (b) density matrix for the steady-state cavity field. It is clearly Wigner-positive. (c) Wigner function and (d) density matrix for the output state with a $T=2.5$ boxcar filter. Negativity appears in the Wigner function, and in the density matrix, the vacuum population is reduced and the two-photon population is increased compared to the cavity field. (e) Wigner function and (f) density matrix for the output field with a $T=5$ boxcar filter, which gives the largest WLN (0.05). Here, the vacuum population is further reduced and the two-photon population is even more prominent.
For clarity in the figure we truncate the density matrices at $n=3$, but the full states occupy a larger Fock space.
}
\label{fig:Wigner-KPO}
\end{figure}

The rather large two-photon populations in the KPO output field may also result in more quantum coherence, namely, in larger density matrix elements $\rho_{02}$ and $\rho_{20}$. This is because the magnitude of these matrix elements is bounded by the populations of the vacuum and two-photon states: $\vert \rho_{0 2} \vert \leq \sqrt{\rho_0 \rho_2}$ (the equality is only achieved for a pure state). Quantum coherence typically translates into negative regions in the Wigner function. Unfortunately there is no simple relation between the $\rho_{02}$ coherence and Wigner negativity, as higher-order Fock states heavily influence the negativity.


In Fig.~\ref{fig:Wigner-KPO} we show the Wigner function and density matrix for the $K=0.5$, $\beta=0.65$ steady-state cavity field, and compare this with the output field for $T=2.5$ and $T=5.0$ boxcar filters. Surprisingly, even though the cavity steady-state field is always Wigner positive~\cite{Kheruntsyan1997, Bartolo2016}, the KPO output field can obtain a negative Wigner function for certain values of the filter width $T$. Negativity in the output increases as a function of $T$ until $T=5.0$ where it peaks, and goes back down if $T$ is further increased. Interestingly, the peak occurs at $T=5$ not only for $\beta=0.65$, but for all $\beta\gtrsim 0.3$. The $\rho_{02}$ coherences are similar for the two output fields, and they are only slightly larger than for the cavity field. But the main contribution to the cavity state coherence is from the vacuum population, which inhibits Wigner negativity.


As the source of quantum coherence is the two-photon drive and the output state is favoured towards even number states, it is no surprise that the KPO output Wigner function resembles a two-component kitten or cat state. Nevertheless, the resulting nonclassical states have a very low purity, roughly 0.617 for the most Wigner-negative state observed.

\begin{figure}[h!]
\centering
\includegraphics[width=\columnwidth]{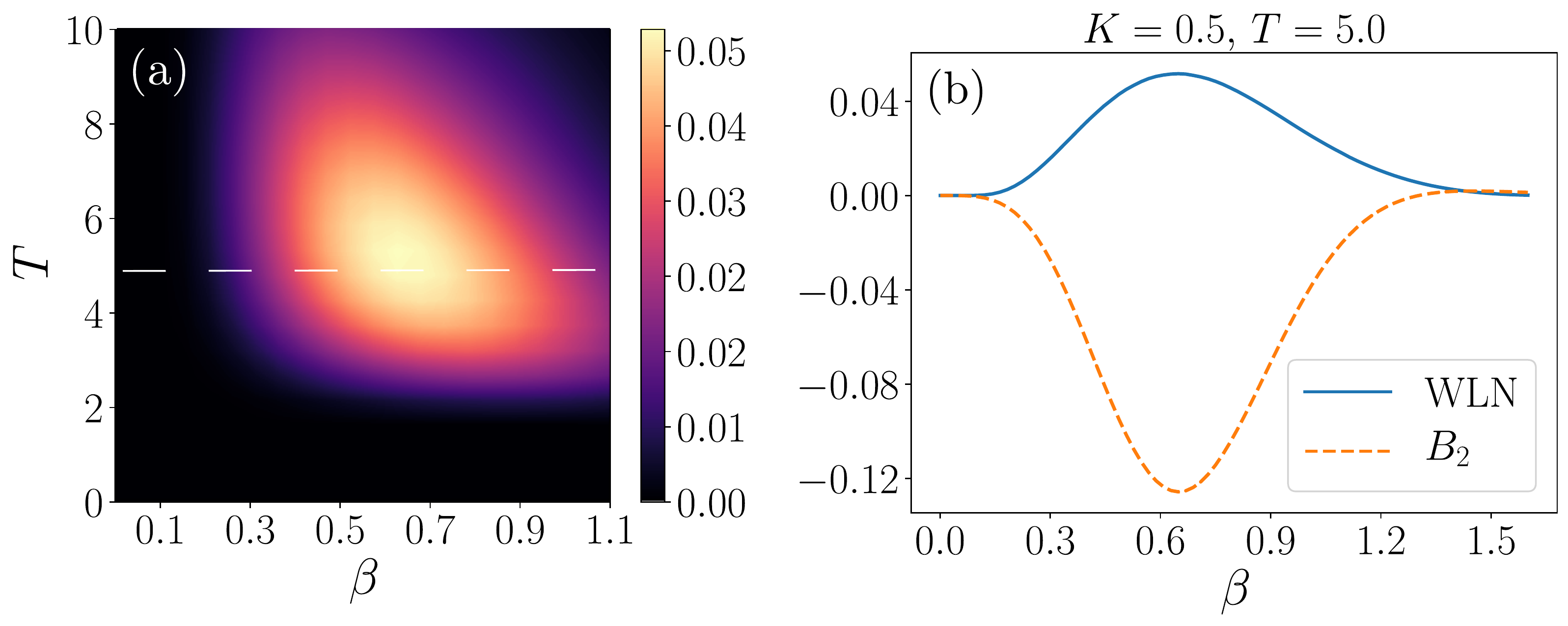}
\caption{(a) Contour map of the WLN as a function of $\beta$ and $T$ for $K=0.5$. The dashed line indicates $T=5$. (b) shows the WLN and Klyshko $B_2$ coefficient as a function of $\beta$ for a fixed $T=5$. The WLN and $B_2$ are clearly related.
}
\label{fig:wigner-klyshko}
\end{figure}

A possible measure of the negativity of the Wigner function is the integrated Wigner negativity~\cite{Kenfack2004},
or alternatively the more recently introduced Wigner Logarithmic Negativity (WLN)~\cite{Albarelli2018}
\begin{equation}
    \wln=\log\left(\int\lvert W(x,p)\rvert \dif x \dif p\right).
\end{equation}
It has the property $\wln>0$ when the Wigner function $W(x,p)$ has a negative part, and has the benefit of being an additive resource monotone~\cite{Chitambar2019Apr}.
In Fig.~\ref{fig:wigner-klyshko}(a) we show a map of the WLN as a function of both the two-photon drive strength $\beta$
and the boxcar width $T$ for $K = 0.5 $.
As can be seen, the maximum value of the negativity is achieved near $T = 5$.
This holds for $K \geq 0.3$.
In the strong nonlinearity regime the maximum negativity occurs for $\beta/K \simeq 1$.

It is interesting to compare the behaviour of the Klyshko coefficient $B_2$ and the WLN. This is shown in Fig.~\ref{fig:wigner-klyshko}(b). As it can be seen, the nonclassical population oscillations around the two-photon state which result in negative values of $B_2$ directly translate into nonclassicality of the Wigner function. We show this correspondence for $T = 5$, but it holds for every value of $T$. 

These results establish that, whereas the cavity decay rate imposes a detection bandwidth for which the cavity state can be output with high fidelity, detection with a wave packet beyond this natural limit may reveal a complete different nature of the output field. In addition, while single-photon losses may destroy quantum coherence inside the cavity, it does not represent a loss mechanism for the output field. In the next section, we will briefly study the effects of the temporal profile of the wave packet.

\subsection{Impact of the filter function on the Wigner negativity}


So far, we have for convenience selected a temporal mode of the cavity output field with a boxcar filter, but any square-integrable function can define a bosonic mode in accordance with Eq.~\eqref{eq:wavepacket_creationop}. Since the filter response can significantly change the nature of the detected state~\cite{Rohde2007Apr}, it is reasonable to suspect that the choice of filter function can have an impact on the observed Wigner negativity. In this section, we are going to target the temporal mode profile which gives the maximum WLN by means of numerical optimization.

\begin{figure}[h!]
\centering
\includegraphics[width=0.8\columnwidth]{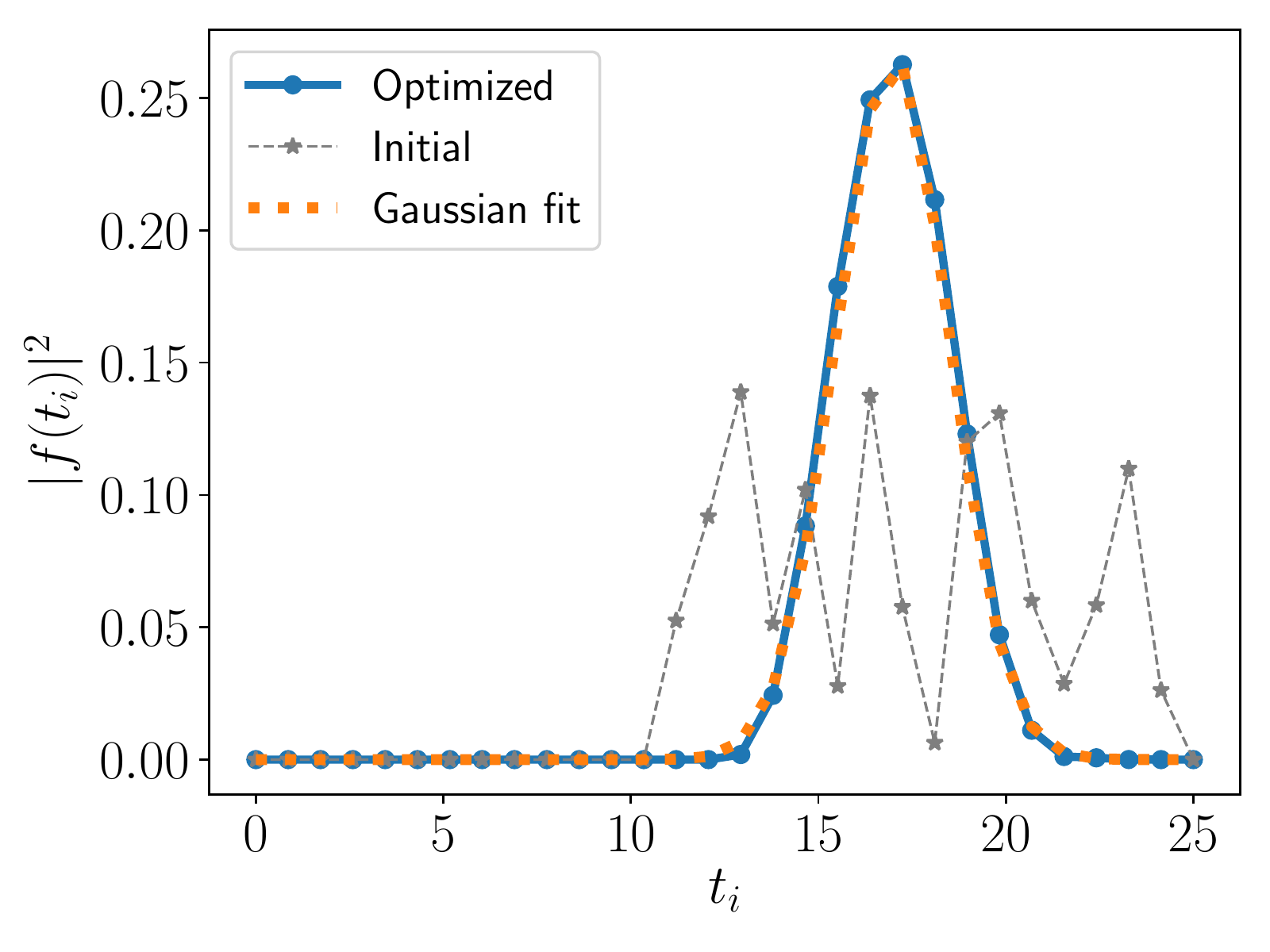}
\caption{Numerical optimization of the filter function at discrete time points $t_i$. The initial filter function
array was constructed by generating random numbers in the interval $[0,1]$ and then normalizing. The filter array was then optimized for maximum WLN. A Gaussian curve can be well fitted to the optimized filter array, here giving $\sigma=2.2$ and $\mu=17.0$. In this example with system parameters $\beta=0.65$ and $K=0.5$, the filter array consists of 18 points and the obtained WLN is 0.1. }
\label{fig:optimized_filter}
\end{figure}

\begin{figure}
    \centering
    \includegraphics[width=0.8\columnwidth]{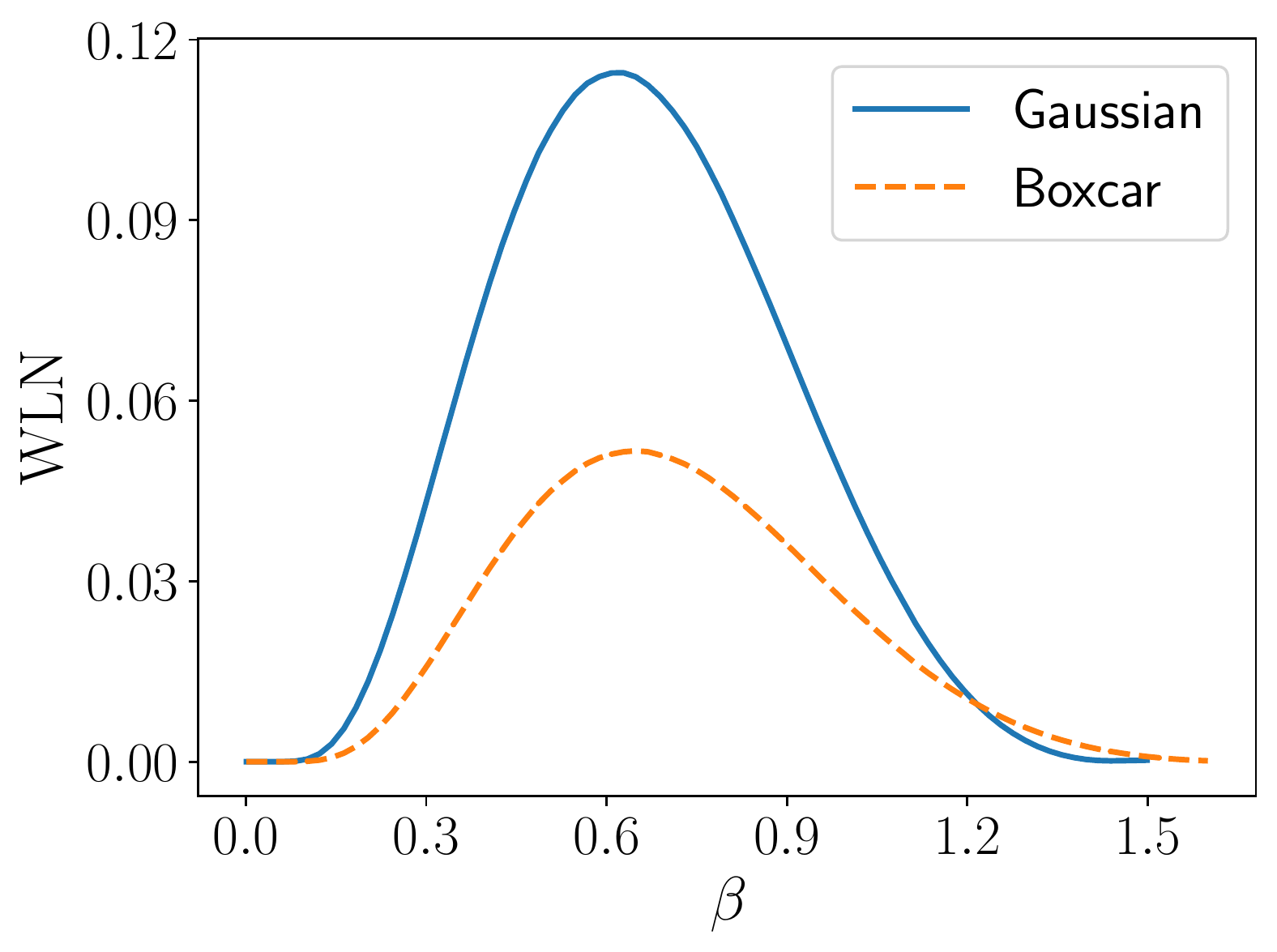}
    \caption{A comparison of the maximum WLN for the boxcar and Gaussian filters, as a function of $\beta$ with the filter widths fixed to the values that give the largest possible WLN: $T=5$ for the boxcar and $\sigma=2.3$ for the Gaussian filter. With the Gaussian filter the maximum WLN is doubled compared to the boxcar.}
    \label{fig:compare_box_gauss}
\end{figure}

Since there is literally an infinite number of possible temporal modes, the best filter could easily be left out if a selection of filter functions were tested manually. To ensure that the best filter function was found, even if it was not a well-behaved, smooth function, we performed an optimization of the numerical array that represents the filter using the \texttt{scipy.optimize} package~\cite{2020SciPy-NMeth}. The part of the filter that is zero until steady-state has been reached was fixed and not included in the optimization. Besides the first and final points being zero, the initial filter was random (but properly normalized, i.e.\ $\sum_i f_i^2=1$). These constraints on the first and final points as well as normalization was enforced during the optimization. Due to the randomness of the initial filter, the results of different optimization runs were not identical. But in general, the optimized filter obtained a Gaussian shape. A representative example is displayed in Fig~\ref{fig:optimized_filter}.
In fact, the maximum WLN obtained with a Gaussian filter is twice the maximum of the boxcar. A comparison between the two is shown in Fig.~\ref{fig:compare_box_gauss} for $T=5$ which gives the maximum WLN for the boxcar filter, and $\sigma=2.3$ which gives the maximum WLN for the Gaussian filter.

%
%

%


\section{Summary \& Conclusions}\label{sec:conclusions}

In this paper we have studied the nonclassicality of the output field of the steady-state Kerr parametric oscillator, in terms of the Wigner function and Klyshko coefficients. Our main result is that whereas the KPO cavity is Wigner-positive in the steady state, the output field can be Wigner-negative, depending on the properties of the selected field mode. In order to obtain the state of the output field we have defined bosonic modes in terms of wave packet functions, utilizing the new "input-output with quantum pulses" formalism introduced by Kiilerich and M{\o}lmer~\cite{Kiilerich2020}, which allows us to obtain the density matrix of the output field.

We also revisited the well-studied linear parametric oscillator. The linear PO is instrumental for the generation of quadrature-squeezed states of light. It is also a paradigmatic example of the different properties exhibited by the cavity and output fields of a continuously driven setup. While the results for the KPO were obtained by numerical simulations, for this linear system we could study the properties of the filtered output field  by reconstructing its density matrix analytically from two-time output field correlations. Here we also explored the so-called even-odd population oscillations as a function of the temporal width of the wave packet function. To the best of our knowledge, these oscillations have previously only been studied by direct photon detection, which is insensitive to the mode structure of the field.

We could then contrast the output of the PO to that of the KPO. The nonlinear Kerr parametric oscillator is also ubiquitous in quantum optics literature. But even though its cavity steady-state has been analytically solved, to the best of our knowledge the output field has not been studied beyond its squeezing properties. We found that the presence of the nonlinearity leads to stronger population oscillations, which is expressed by the larger magnitude of the Klyshko coefficient. This is what gives rise to the Wigner negativity in the KPO output, as the magnitude of the Klyshko coefficient directly correlates to the integrated Wigner logarithmic negativity (WLN). Furthermore, by numerical optimization we have verified that the nonclassical properties of the output field are dependent on the chosen wave packet function, and that a Gaussian wave packet maximizes the WLN.


Typically, the important parameter responsible for driven nonlinear oscillators to reach quantum regimes is the ratio between the Kerr parameter and cavity decay rate, i.e., efficiency of quantum nonlinear effects requires a high nonlinearity with respect to dissipation~\cite{Gevorgyan2013Jun}. In contrast, there is no need for a strong nonlinearity to observe Wigner negativity in the KPO steady-state output, as $K/\gamma = 0.7$ is sufficient to observe the largest Klyshko negativity and the largest Wigner negativity.

Our results could realistically be verified experimentally. Notably, in superconducting circuit experiments the nonlinearity strength can be tuned~\cite{Bell2012Sep}. Quantum state tomography of propagating microwave fields has been established using both phase insensitive~\cite{Eichler2011, Fedorov2016} and sensitive amplification~\cite{Mallet2011}, and more recently via parity detection~\cite{Besse2019}. In addition, the Klyshko nonclassicality criterion is amenable to be tested for propagating fields following a recent proposal for a microwave number-resolved photon counter~\cite{Dassonneville2020}.
Finally, our approach could be extended to study the output field properties of recent higher-order squeezing realizations~\cite{Sandbo2020, Svensson2018}. Also, it would be interesting to study the role of filtering in the output entanglement properties of multimode setups.






\section{Acknowledgments}

The authors would like to thank Klaus M{\o}lmer and Chris Wilson for valuable discussions.
IS acknowledges support from Chalmers Excellence Initiative Nano.
FQ and GJ acknowledge the financial support from the Knut and Alice Wallenberg Foundation
through the Wallenberg Center for Quantum Technology (WACQT).

\begin{appendix}

\section{Density matrix from the covariance matrix for a Gaussian state}\label{app:sect-analytical}

A Gaussian state is defined by its covariance matrix $V$ with
matrix elements:
\begin{align}
V_{11} &= \langle \hat x^2 \rangle \\
V_{22} &= \langle \hat p^2 \rangle \\
V_{12} &= V_{21} = \frac{1}{2}  \langle \hat x \hat p + \hat p \hat x \rangle ,
\end{align}
with $\hat x$ and $\hat p$ the position and momentum quadratures respectively
\begin{align}
\hat x &= \frac{1}{\sqrt{2}} (  \hat b^\dagger + \hat b   ) \\
\hat p &= \frac{i}{\sqrt{2}} (  \hat b^\dagger - \hat b  ) ,
\end{align}
with $\hat b$ ($\hat b^\dagger$) the bosonic annihilation (creation) operation which might refer to
a cavity or filtered propagating mode.
Here we are assuming that $\langle \hat x \rangle = \langle \hat p \rangle = 0$ which is true for the models
studied in the main text.
Recall that in the steady-state of \eqref{eq:qme} with
Hamiltonian \eqref{eq:Hamiltonian} we have
$\langle \hat c \rangle_{\rm ss} = 0$
and consequently, $\langle \hat A_f \rangle = 0$
for the filtered output field

The density matrix elements in the Fock or number basis
can be recovered from the covariance matrix $V$ by means of the relation
\begin{equation}\label{app:rho}
\langle m \vert \rho \vert n \rangle = \left( d + \frac{t}{2} + \frac{1}{4} \right)^{-1/2} \frac{1}{\sqrt{m! \,n! }} H_{mn}^{\{ R\}} (0,0) ,
\end{equation}
with $d = \det V$ and $t = {\rm tr}\,V$ the determinant and the trace of the covariance matrix, respectively~\cite{Dodonov1984}. The so-called  multidimensional Hermite polynomials  $H_{mn}^{\rm R} (x_1,x_2) $
are defined in terms of a $2 \times 2$ matrix $R$ with elements~\cite{Dodonov1994}
\begin{align}
R_{11} &= \left(  2d + t + \frac{1}{2}  \right)^{-1} (V_{11} - V_{22} - 2 \imi V_{12}) \label{app:R11}\\
R_{22} &= \left(  2d + t + \frac{1}{2}  \right)^{-1} (V_{11} - V_{22} + 2 \imi V_{12}) \\
R_{12} &= R_{21} = \left(  2d + t + \frac{1}{2}  \right)^{-1} \left( \frac{1}{2} - 2 d \right) \label{app:R12} .
\end{align}
The arguments $x_1$ and $x_2$ are related to the first moments of the field, which in our case are always zero.
We get
\begin{align}\label{app:H-matrix}
H_{mn}^{\{ R\}} (0,0) = m! n! \sqrt{\frac{R_{11}^m R_{22}^n}{2^{m+n}}} \sum_{k=0}^{{\rm min}(m,n)}
\left( \frac{- 2 R_{12} }{\sqrt{R_{11} R_{22}} } \right)^k \times \nonumber\\
\times \frac{1}{k! (m-k)! (n-k)!} H_{m-k} \left( 0 \right) H_{n-k} \left(  0 \right) ,
\end{align}
with $H_n(x)$ being the $n$th order Hermite polynomial.


\section{Suppression of odd Fock states populations}\label{app:sect-odd-Fock}

The Heisenberg uncertainty principle puts a lower bound on the minimum value of $d = \det V$, where $V$ is the covariance matrix of a quantum state, defined by Eqs.~\eqref{app:R11}-\eqref{app:R12}.
Under the quadrature normalization used here we have $d \geq 1/4$.
A \emph{minimum uncertainty state} is by definition
a state for which $d = 1/4$. Examples of these
are coherent states and the so-called
\emph{ideal} squeezed states~\cite{Gerry}.
In a squeezed state, noise (the variance) in one quadrature is reduced below the vacuum level at the expense of increased noise in the orthogonal quadrature. For an ideal squeezed state, the product of the quadrature variances equals the lower bound.

A quintessential example of an ideal squeezed state is the  squeezed vacuum state, that is, the state that results from the action of the unitary squeezing operator
\begin{equation}
    \hat S (\xi) = \mathrm e^{(\xi^* \hat b^2 - \xi \hat b^{\dagger 2} )/2},
\end{equation}
with $\xi = r \exp(i \theta)$ ($r, \theta \in \mathbb{R}$) on the photon vacuum state $\vert 0 \rangle$.
For $\theta = 0$, the variances satisfy
\begin{equation}
V_{11} = \langle \hat x^2 \rangle = \frac{1}{2} \exp(-2 r),
\end{equation}
and
\begin{equation}
 V_{22}= \langle \hat p^2 \rangle = \frac{1}{2} \exp(+2 r),
\end{equation}
and thus, $d = 1/4$~\cite{Gerry}.

The suppression of the odd Fock state populations is a consequence of ideal squeezing.
Indeed, the condition $d = 1/4$ results in $R_{12} = 0$ in Eq.~\eqref{app:R12}. If this is the case, only the term $k=0$ will contribute to the summation in Eq.~\eqref{app:H-matrix}. For $m = n$ the latter reduces to $(H_{m} (0)/m!)^2$. All of the odd-order Hermite polynomials are identical to zero at the origin ($x=0$).
Consequently, $\braket{n|\rho| n} = 0$ in Eq.~\eqref{app:rho} for odd $n$.

\section{Time-filtering of a coherent state}\label{app:sect-coherent}

Let us assume that the steady-state of the cavity field is a coherent state.
A coherent state is completely defined by the first moment of the bosonic field operator, i.e., $\langle \hat c \rangle_{\rm ss}$
with higher-order moments factorizing
in terms of it.
Following the input-output relation, we have for the filtered output field moments:
\begin{align}
\langle \hat A_f \rangle &= \sqrt{\gamma} \int_0^\infty {\rm d}t \, f(t) \langle \hat c(t) \rangle_{\rm ss}, \\
\langle \hat A_f^\dagger \hat A_f \rangle &= \gamma \int_0^\infty  {\rm d}t \, {\rm d}t' \, f(t) f(t')  \langle \hat c(t)^\dagger \hat c(t') \rangle_{\rm ss},
\end{align}
where for simplicity we are assuming the filter function $f$ to be real.
Using the coherent field property that higher order moments factorize, we get
\begin{align}
\langle \hat A_f \rangle &= \sqrt{\gamma}  \langle \hat c \rangle_{\rm ss} \int_0^\infty {\rm d}t \, f(t), \\
\langle \hat A_f^\dagger \hat A_f \rangle &= \gamma
\vert  \langle \hat c \rangle_{\rm ss} \vert^2 \left( \int_0^\infty {\rm d}t \, f(t) \right)^2 ,
\end{align}
with similar relations holding for third- and higher-order correlations.
For a boxcar and Gaussian wave packets, the above time integral depends on the width and the variance ($T$ and $\sigma$) of these functions respectively. That is, the effect of the filter is just to rescale the cavity moments proportionally to the temporal width of the filter.
Finally, for an incoherent superposition of coherent states the above arguments hold for every coherent state in the mixture.


\end{appendix}

\bibliography{references}

\end{document}